\documentclass[draft]{agujournal2018}
\usepackage{apacite}
\usepackage{natbib}
\usepackage{amssymb, amsmath}
\usepackage{url} %this package should fix any errors with URLs in refs.
\usepackage[normalem]{ulem}
\usepackage[final]{pdfpages}

\draftfalse

\journalname{Geophysical Research Letters}

\begin{document}

%% ------------------------------------------------------------------------ %%
%  Title
%
% (A title should be specific, informative, and brief. Use
% abbreviations only if they are defined in the abstract. Titles that
% start with general keywords then specific terms are optimized in
% searches)
%
%% ------------------------------------------------------------------------ %%

% Example: \title{This is a test title}

\title{Ocean dynamics of outer solar system satellites}

%% ------------------------------------------------------------------------ %%
%
%  AUTHORS AND AFFILIATIONS
%
%% ------------------------------------------------------------------------ %%

% Authors are individuals who have significantly contributed to the
% research and preparation of the article. Group authors are allowed, if
% each author in the group is separately identified in an appendix.)

% List authors by first name or initial followed by last name and
% separated by commas. Use \affil{} to number affiliations, and
% \thanks{} for author notes.
% Additional author notes should be indicated with \thanks{} (for
% example, for current addresses).

% Example: \authors{A. B. Author\affil{1}\thanks{Current address, Antartica}, B. C. Author\affil{2,3}, and D. E.
% Author\affil{3,4}\thanks{Also funded by Monsanto.}}

\authors{K. M. Soderlund\affil{1}}

% \affiliation{1}{First Affiliation}
% \affiliation{2}{Second Affiliation}
% \affiliation{3}{Third Affiliation}
% \affiliation{4}{Fourth Affiliation}

\affiliation{1}{Institute for Geophysics, Jackson School of Geosciences, The University of Texas at Austin, Austin, Texas, USA}
%(repeat as many times as is necessary)

%% Corresponding Author:
% Corresponding author mailing address and e-mail address:

% (include name and email addresses of the corresponding author.  More
% than one corresponding author is allowed in this LaTeX file and for
% publication; but only one corresponding author is allowed in our
% editorial system.)

% Example: \correspondingauthor{First and Last Name}{email@address.edu}

\correspondingauthor{Krista Soderlund}{krista@ig.utexas.edu}

%% Keypoints, final entry on title page.

% Example:
% \begin{keypoints}
% \item	List up to three key points (at least one is required)
% \item	Key Points summarize the main points and conclusions of the article
% \item	Each must be 100 characters or less with no special characters or punctuation
% \end{keypoints}

%  List up to three key points (at least one is required)
%  Key Points summarize the main points and conclusions of the article
%  Each must be 100 characters or less with no special characters or punctuation

\begin{keypoints}
\item Ocean dynamics are important for the habitability of icy ocean worlds.
\item Strong ocean currents likely exist in Enceladus, Titan, Europa, and Ganymede.
\item Convective heat transfer in the ocean is predicted to vary with latitude, which would modify the thermophysical structure of the ice shell.
\end{keypoints}

%% ------------------------------------------------------------------------ %%
%
%  ABSTRACT
%
% A good abstract will begin with a short description of the problem
% being addressed, briefly describe the new data or analyses, then
% briefly states the main conclusion(s) and how they are supported and
% uncertainties.
%% ------------------------------------------------------------------------ %%

%% \begin{abstract} starts the second page

\begin{abstract}
Ocean worlds are prevalent in the solar system. Focusing on Enceladus, Titan, Europa, and Ganymede, I use rotating convection theory and numerical simulations to predict ocean currents and the potential for ice-ocean coupling. When the influence of rotation is relatively strong, the oceans have multiple zonal jets, axial convective motions, and most efficient heat transfer at high latitudes. This regime is most relevant to Enceladus and possibly to Titan, and may help explain their long-wavelength topography. For a more moderate rotational influence, fewer zonal jets form, Hadley-like circulation cells develop, and heat flux peaks near the equator. This regime is predicted for Europa, where it may help drive geologic activity via thermocompositional diapirism in the ice shell, and is possible for Titan. Weak rotational influence allows concentric zonal flows and overturning cells with no preferred orientation. Predictions for Ganymede's ocean span multiple regimes. \\
{\bf Plain Language Summary:} The outer solar system is host to a large number of diverse satellites, many of which likely have global oceans beneath their outer icy shells. I use theoretical arguments and numerical models to make predictions about ocean currents and heat transfer across such oceans. Our results suggest that strong ocean currents exist in Enceladus, Titan, Europa, and Ganymede, and cause the transfer of heat to vary with latitude that may modify the overlying ice shell.
\end{abstract}

%% ------------------------------------------------------------------------ %%
%
%  TEXT
%
%% ------------------------------------------------------------------------ %%

%%% Suggested section heads:
% \section{Introduction}
%
% The main text should start with an introduction. Except for short
% manuscripts (such as comments and replies), the text should be divided
% into sections, each with its own heading.

% Headings should be sentence fragments and do not begin with a
% lowercase letter or number. Examples of good headings are:

% \section{Materials and Methods}
% Here is text on Materials and Methods.
%
% \subsection{A descriptive heading about methods}
% More about Methods.
%
% \section{Data} (Or section title might be a descriptive heading about data)
%
% \section{Results} (Or section title might be a descriptive heading about the
% results)
%
% \section{Conclusions}

\section{Introduction}

Exploration of the outer solar system has shown that subsurface oceans may be relatively common in the interiors of icy satellites and dwarf planets \citep{NimmoPappalardo16,Lunine17}. Strong evidence exists for oceans in the Saturnian satellites Enceladus and Titan, with oceans also potentially present in Mimas and Dione. In the Jovian system, there is compelling evidence for a Europan ocean and oceans are predicted within Ganymede and potentially Callisto as well \citep[c.f.][]{HartkornSaur17}. Kuiper belt objects, such as Pluto, Charon, and the Neptunian satellite Triton, may also have subsurface oceans.

The presence of liquid water makes these ocean worlds compelling astrobiological targets. However, the dynamics of these oceans also play a role in promoting habitable environments. For example, heat and material exchange between the seafloor and ice shell will be enhanced if the ocean is unstable to convection \citep[e.g.,][]{VanceGoodman09,SoderlundEA14}, has mechanically driven flows \citep[e.g.,][]{Tyler08,LemasquerierEA17,WilsonKerswell18,RoviraNavarroEA19}, or has fluid motions driven by magnetic pumping \citep[][]{GissingerPetitdemange19}. Currents and turbulence tend to mix the ocean waters, which implies the presence of strong thermal and compositional gradients near the top and bottom of the ocean that life may take advantage of. Mixing efficiency may vary spatially, so these motions are also important for the distribution of potential bionutrients. In addition, heat transfer from the ocean will influence where the ice shell melts and freezes. Melting of the ice shell and freezing of the ocean will impact the salt budget, especially near the ice-ocean interface, a habitable environment in analogous terrestrial ice shelves \citep[e.g.,][]{DalyEA13}. Moreover, accreted ice depleted in salts may have positive buoyancy and lead to upwelling thermocompositional diapirs in the ice shell that would link the surface and subsurface \citep[e.g.,][]{PappalardoBarr04,SoderlundEA14}.

Here, I focus on the ocean dynamics of Enceladus and Titan given the abundance of data from the {\it Cassini} mission and of Europa and Ganymede in preparation for the upcoming {\it Europa Clipper} \citep{PhillipsPappalardo14} and {\it JUICE} \citep{GrassetEA13} missions. Scaling laws for rotating convection systems are used to make predictions about their convective behaviors in Section 2, and numerical models of global ocean convection characterizing the predicted regimes are presented in Section 3. Implications for icy satellites are explored in Section 4, and the challenges of extrapolating to realistic ocean conditions are discussed in Section 5.

\section{Rotating Convection Scaling Laws}

Convection characteristics depend critically on the relative importance of rotation, which tends to organize the fluid into columns aligned with the rotation axis, increase the critical Rayleigh number, constrain heat transfer efficiency, and drive zonal flows \citep[e.g.,][]{AurnouEA15}. \citet{ChengEA18} combines asymptotic predictions, laboratory experiments, and numerical simulations to review the behavior of rotating thermal convection as a function of the dimensionless Ekman, Rayleigh, and Prandtl numbers. The Ekman number, $E=\nu/2\Omega D^2$, represents the ratio of rotational to viscous timescales; thus, low $E$ signifies rapid rotation rates in planetary interiors. The Rayleigh number, $Ra = \alpha g \Delta T D^3 / \nu \kappa$, is the ratio of the thermal diffusion time to the viscous buoyant rise time; large $Ra$ denotes strong buoyancy forcing. The Prandtl number, $Pr = \nu / \kappa$, defines the ratio of thermal to viscous diffusion timescales. Here, $\nu$ is kinematic viscosity, $\Omega$ is rotation rate, $D$ is ocean thickness, $\alpha$ is thermal expansivity, $g$ is gravitational acceleration, $\Delta T$ is superadiabatic temperature contrast, and $\kappa$ is thermal diffusivity.

\begin{figure}[ht]
\centering
\includegraphics[width=34pc]{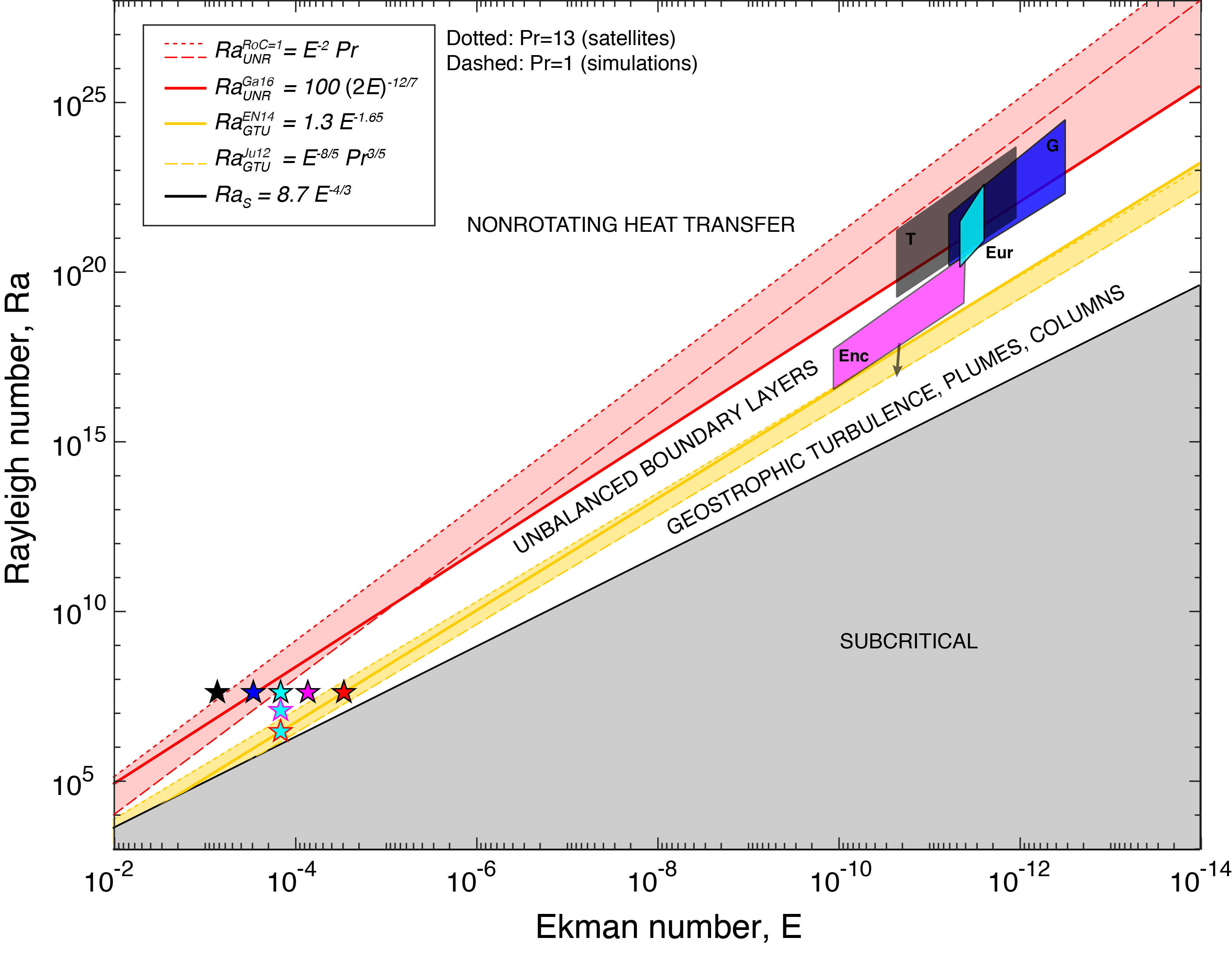}
\caption{Convective regime diagram following \citet{ChengEA18} with superimposed parameter estimates for Enceladus (magenta), Titan (gray), Europa (cyan), and Ganymede (blue) superimposed (see Table~\ref{tab:physparam}). Our numerical simulations are denoted by stars, where interior color denotes Ekman number and exterior color denotes Rayleigh number. Interior: red ($E=3.0 \times 10^{-5}$), magenta ($E=7.5 \times 10^{-5}$), cyan ($E=1.5 \times 10^{-4}$), blue ($E=3.0 \times 10^{-4}$), and black ($E=7.5 \times 10^{-4}$). Exterior: black ($Ra=3.4 \times 10^7$), magenta ($Ra=1.0 \times 10^7$), and red ($Ra=2.4 \times 10^6$). The black line denotes the scaling for the onset of convection at Rayleigh number $Ra_S$ \citep{Chandrasekhar61}, the yellow lines bound the range of predicted transitions from the GT regime to the UBL regime that occurs at a Rayleigh number $Ra_{GTU}$ \citep{EckeNiemela14,JulienEA12_PRL}, and the red lines bound the range of predicted transitions from the UBL regime to the NR regime that occurs at a Rayleigh number $Ra_{UNR}$ \citep{GastineEA16,Gilman77}. The $Ra^{RoC=1}_{UNR}$ and $Ra^{EN14}_{GTU}$ scalings depend on the Prandtl number; dotted lines assume $Pr=13$ following the upper estimate for icy satellite oceans while dashed lines assume $Pr=1$ as used in the simulations.}
\label{fig:regimes}
\end{figure}

\citet{ChengEA18} identify five rotating convection regimes: columnar, plumes, geostrophic turbulence (GT), unbalanced boundary layer (UBL), and nonrotating heat transfer (NR) (see Fig.~\ref{fig:regimes}). 
Near onset, convection in the bulk fluid manifests as Taylor columns aligned with the rotation axis (``columnar" regime).  With increased buoyancy forcing, the columns begin to deteriorate such that they no longer extend fully across the fluid layer (``plumes" regime). Convection eventually becomes vigorous enough for strong mixing in the bulk fluid  (``geostrophic turbulence" regime). Despite the disappearance of coherent vertical structures, the Coriolis force still imposes a vertical stiffness on the flow field. These regimes are shown collectively on Figure~\ref{fig:regimes}. The influence of rotation is lost locally at Rayleigh numbers exceeding $Ra_{GTU}$, which corresponds to the breakdown of geostrophy (balance between Coriolis and pressure gradient forces) in the thermal boundary layers (``unbalanced boundary layers" regime). For Rayleigh numbers greater than $Ra_{UNR}$, the influence of rotation is lost globally (``nonrotating heat transfer" regime). As reviewed by \citet{ChengEA18}, significant debate exists in the community on the scaling laws for $Ra_{GTU}$ and $Ra_{UNR}$. Rather than assume a single scaling law for each transition, I consider upper and lower bound scaling laws for each regime and highlight the resulting range of parameter space for each regime transition in Figure~\ref{fig:regimes}.

Using this regime diagram, one can predict the convective regime of a system if the Ekman, Rayleigh, and Prandtl numbers can be estimated (see Table~\ref{tab:physparam}). The Prandtl number depends only on fluid properties and is estimated to be $Pr \sim 10$ for the satellite oceans \citep{AbramsonEA01,NayarEA16}. The Ekman number is also relatively easy to calculate since it only requires assumptions about the fluid viscosity, rotation rate, and ocean thickness. I use the internal structure models of \citet{VanceEA18} (see their Tables 5-8) to obtain ocean thicknesses $D_{ocean}$ for six combinations of possible outer ice shell thicknesses and ocean compositions for each satellite. Enceladus' ocean has the largest Ekman number of $E \sim \mathcal{O}(10^{-10})$, while the ocean of Ganymede has the lowest at $E \sim \mathcal{O}(10^{-13})$.

The Rayleigh number is more difficult to estimate because it requires knowledge of the superadiabatic temperature contrast $\Delta T$. One can derive an estimate, however, using the relationship between the Rayleigh number and the convective heat transfer efficiency as measured by the Nusselt number, $Nu = q D / \rho C_p \kappa \Delta T$. Following \citet{SoderlundEA14}, I leverage $Nu-Ra$ scalings to solve for $\Delta T$ algebraically and consider both non-rotating and rapidly rotating scaling laws to give end-member estimates. More recent scaling laws for rotating spherical shells are used here, however. In the non-rotating regime, heat transfer is expected to be independent of the Ekman number and follow the theoretical limit of $Nu = 0.07 Ra^{1/3}$ \citep[e.g.,][]{GastineEA15}. Conversely, in the rapidly rotating limit, heat transfer is predicted to follow $Nu = 0.15 Ra^{3/2} (2E)^2$ \citep{GastineEA16}. As a result, the temperature contrast is given by
\begin{equation}
\Delta T = 7.3 \left (\frac{\nu}{\alpha g \rho C_p} \right )^{1/4} q^{3/4}
\end{equation}
in the non-rotating regime and by 
\begin{equation}
\Delta T = 2.1 \left ( \frac{\Omega^4 \kappa}{\rho^2 C_p^2 \nu \alpha^3 g^3} \right )^{1/5} (q^2 D)^{1/5}
\end{equation}
in the rapidly-rotating regime. 

Here, I assume the heat flux $q$ from each of the six interior models per satellite, noting that the lower $q$ estimates are associated with thicker ice Ih shells \citep{VanceEA18}. Although these values are generally consistent with the literature, the minimum heat fluxes tend to exceed those predicted for radiogenic heating in the mantle at present day \citep[e.g.,][]{BlandEA09} and the upper bound for Ganymede is appropriate for a past active period \citep[e.g.,][]{DombardMcKinnon01}. If $q$ is decreased by an order of magnitude, the lower bounds for $\Delta T$, and therefore $Ra$, only decrease by a factor of 2.5 per eqn. (2). Our global estimates also neglect spatial variations in heat flow that may be locally strong at Enceladus \citep{ChobletEA17}, for example, where narrow mantle upwellings can reach $1-5$ W/m$^2$ (the global average, however, is in line with \cite{VanceEA18}). If $q$ is increased by an order of magnitude, the $Ra$ upper bounds increase by a factor 5.6 per eqn. (3). As shown in Table~\ref{tab:physparam}, Rayleigh numbers span from $Ra \sim \mathcal{O}(10^{16})$ for the lower Enceladus limit to $Ra \sim \mathcal{O}(10^{24})$ for the upper Ganymede limit. An important caveat to note here, however, is that these estimates do not include compositional contributions due to salinity gradients. This simplification may be especially significant for Titan since the ocean is hypothesized to have a high concentration of dissolved salts \citep[][]{BalandEA14, MitriEA14}. 

Figures~\ref{fig:regimes} and S1 plot the resulting estimates of the Ekman and Rayleigh numbers on the convective regime diagram. The oceans of Titan, Europa, and Ganymede are predicted to behave similarly since their estimated parameter spaces have considerable overlap. Since these estimates fall near the lower boundary between the UBL and NR regimes, I hypothesize that rotational effects do not dominate the turbulent local-scale convective flows. Conversely, rotation likely has a stronger influence on the ocean of Enceladus, which is also predicted to be primarily in the UBL regime, although extending into the GT transition.

\section{Numerical Convection Models}

Numerical models of global ocean convection are next used to characterize the currents and heat flow patterns. I utilized the pseudospectral code MagIC, version 5.6 \citep[e.g.,][]{Wicht02,GastineWicht12} to simulate 3D, time-dependent, thermal convection of a Boussinesq fluid in a rotating spherical shell with  geometry characterized by the ratio of inner to outer shell radii, $\chi = r_i/r_o = 0.9$. The system is further defined by the Ekman, Rayleigh, and Prandtl numbers. Following \citet{SoderlundEA14}, the boundaries are impenetrable, stress-free, and isothermal. Compositional buoyancy, spatial variations in mantle heat flow, and mechanically driven flows are neglected for simplicity.

Seven models that span a convective regime space consistent with the icy satellite ocean predictions are considered (Fig.~\ref{fig:regimes}). In the first series, the Rayleigh and Prandtl numbers are fixed to $Ra = 3.4 \times 10^7$ and $Pr=1$, and the Ekman number is increased from $E = 3.0 \times 10^{-5}$ to $E = 7.5 \times 10^{-4}$. The second series of models increase the Rayleigh number from $Ra = 2.4 \times 10^6$ to $Ra = 3.4 \times 10^7$ for fixed $E=1.5 \times 10^{-4}$ and $Pr=1$; higher $Ra$ values were not pursued to due computational limitations. Hyperdiffusivities are not employed \citep[c.f.][]{ZhangSchubert00}. The numerical grids have 73 radial points, 320 latitudinal points, and 640 longitudinal points for cases with $E \geq 7.5 \times 10^{-5}$ and 65 radial points, 640 latitudinal points, and 1280 longitudinal points for the $E=3.0 \times 10^{-5}$ case. Each model was initiated with a random temperature perturbation or restarted from a lower $E$ or $Ra$ case. 

Figure~\ref{fig:Vel} shows the mean velocity and temperature fields of each model across the $E$ parameter sweep, while Figure~\ref{fig:HF} shows the normalized heat flux along the outer boundary. In the highest Ekman number case (Fig.~\ref{fig:Vel}a), the zonal and radial flows have comparable magnitudes reminiscent of non-rotating convection. The radial flows have no preferred spatial orientation, while the zonal flows are concentric due to viscous transport of angular momentum \citep{BrunPalacios09}. Ocean temperatures are nearly isothermal away from the boundaries, leading to localized heat flux perturbations along the ice-ocean interface (Fig.~\ref{fig:HF}a).

When the Ekman number is decreased (Fig.~\ref{fig:Vel}b-c), homogenization of absolute angular momentum leads to zonal flows that are retrograde (westward) at large cylindrical radii and prograde (eastward) closer to the rotation axis \citep[e.g.,][]{Gilman78,AurnouEA07,GastineEA13}. The mean radial flows become more organized with a pronounced upwelling near the equator and downwellings at mid-latitudes, essentially forming Hadley-like meridional circulation cells in each hemisphere. Upon further decreasing of the Ekman number (Fig.~\ref{fig:Vel}d), multiple zonal jets that alternate in direction develop, and the mean radial flows retain an equatorial upwelling that becomes more aligned with the rotation axis. Both mean zonal and radial flow speeds decrease by a factor of five compared to the $E=[3.0, 1.5] \times 10^{-4}$ cases. In all three of these models (Fig.~\ref{fig:Vel}b-d), ocean temperatures are characterized by thin thermal boundary layers and warmer equatorial waters. Heat flux peaks at low latitudes (with minima at mid-latitudes) due to the mean overturning circulations with secondary peaks forming at high latitudes due to turbulent heat transfer associated with vertically ascending plumes (Fig.~\ref{fig:HF}b-d).

In the lowest Ekman number case (Fig.~\ref{fig:Vel}e), Coriolis forces organize the flow into narrow structures that are aligned with the rotation axis. Reynolds stresses associated with these columns drive prograde equatorial flow with jets that alternate in direction at higher latitudes due to correlation locally between the azimuthal and cylindrically radial flow components \citep[e.g.,][]{AurnouOlson01,Christensen01,HeimpelEA05,GastineEA14}. Ocean temperatures are not well-mixed, especially at low latitudes, due to the axialized convective flows and strong equatorial jet \citep[e.g.,][]{AurnouEA08}. Consequently, heat flow along the ice-ocean interface peaks at high latitudes with minima near the equator (Fig.~\ref{fig:HF}e). 

A similar trend from three-jet zonal flows, equatorial upwelling, and peak low latitude heat flux to multiple zonal jets, axialized convective flows, and peak high latitude heat flux is found as $Ra$ is decreased (Figs.~S2 and S3).

\section{Implications for Icy Satellites}

In order to apply these models to icy satellite oceans, I assume that the velocity and temperature patterns extrapolate to more extreme parameters following the relative distance between regime boundaries (Fig.~S1). Enceladus' ocean may then be represented by the $E=[3.0 \times 10^{-5}, 7.5 \times 10^{-5}]$ models since both fall approximately between the GT-UBL regime transition and the lower bound of the UBL-NR transition. In contrast, the oceans of Europa, Ganymede, and Titan depend on the UBL-NR transition scaling used. If $Ra^{RoC=1}_{UNR}=E^{-2}Pr$ \citep[e.g.,][]{Gilman77} is assumed, then all of these oceans are near the center of UBL regime such that the $E=[7.5 \times 10^{-5},1.5 \times 10^{-4}]$ models would be most appropriate for these satellites. If the transition instead follows $Ra^{Ga16}_{UNR}=100(2E)^{-12/7}$ \citep[][]{GastineEA16}, then the $E=3.0 \times 10^{-4}$ model would best characterize Europa and the $E=[3.0 \times 10^{-4}, 7.5 \times 10^{-4}]$ models would be most pertinent to Titan and Ganymede.

Below, I discuss the implications for each satellite. Regions with high heat flow are presumed to undergo enhanced melting, leading to ice shell thickness variations. However, large thickness disparities can set up a phenomena known as an ice pump \citep[e.g.,][]{LewisPerkin86} where pressure-induced melting occurs where the ice shell is thick and re-accretes where the ice shell is thin, effectively reducing topography along the base of the ice shell. Since the accretion process is very efficient at excluding impurities in low temperature environments \citep[e.g.,][]{MooreEA94,EickenEA94}, this marine ice may be salt-depleted compared to the overlying ice.  The ice may, therefore, have positive buoyancy due to the associated thermal and compositional density anomalies and rise toward the surface in the form of convective diapirs \citep[e.g.,][]{PappalardoBarr04,SoderlundEA14}. Alternatively, if the ice pump mechanism is not efficient, the ice shell may be more unstable to convection where it is relatively thick \citep{TravisEA12,Goodman14}.

\begin{figure}[ht]
\centering
\includegraphics[width=34pc]{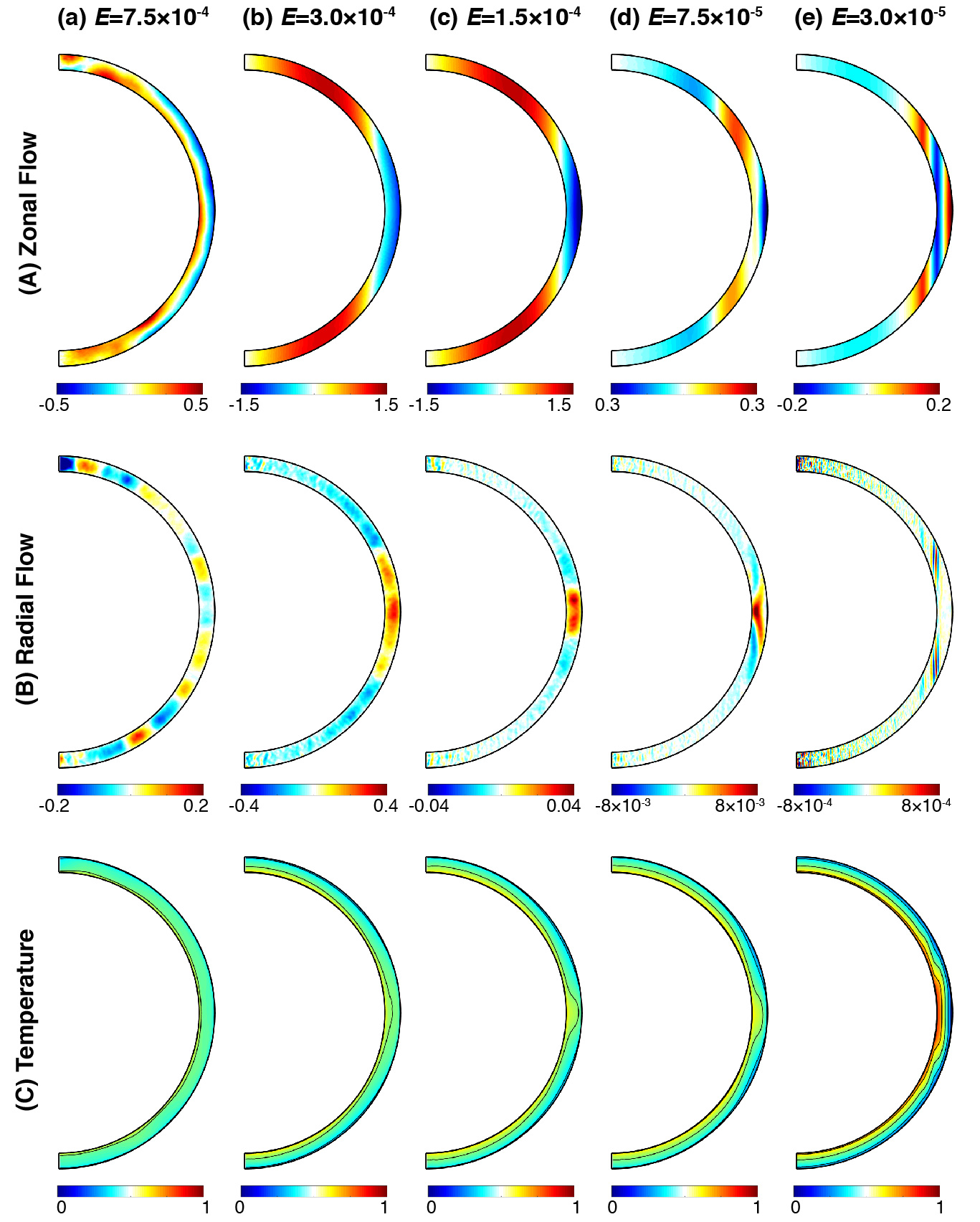}
\caption{Mean velocity and temperature fields averaged over all longitudes and at least 0.02 viscous diffusion times for each model in the Ekman number suite. (A) Zonal flows given in dimensionless Rossby number units, $Ro=U/\Omega D$, which characterizes the ratio of rotational to inertial timescales. Red (blue) denotes prograde (retrograde) currents. (B) Radial flows given in $Ro$ units. Red (blue) denotes upwelling (downwelling) currents. (C) Superadiabatic temperature in dimensionless units with isotherm contours superimposed (interval of 0.2). Red (blue) denotes warm (cool) fluid.}
\label{fig:Vel}
\end{figure}

\begin{figure}[ht]
\centering
\includegraphics[width=34pc]{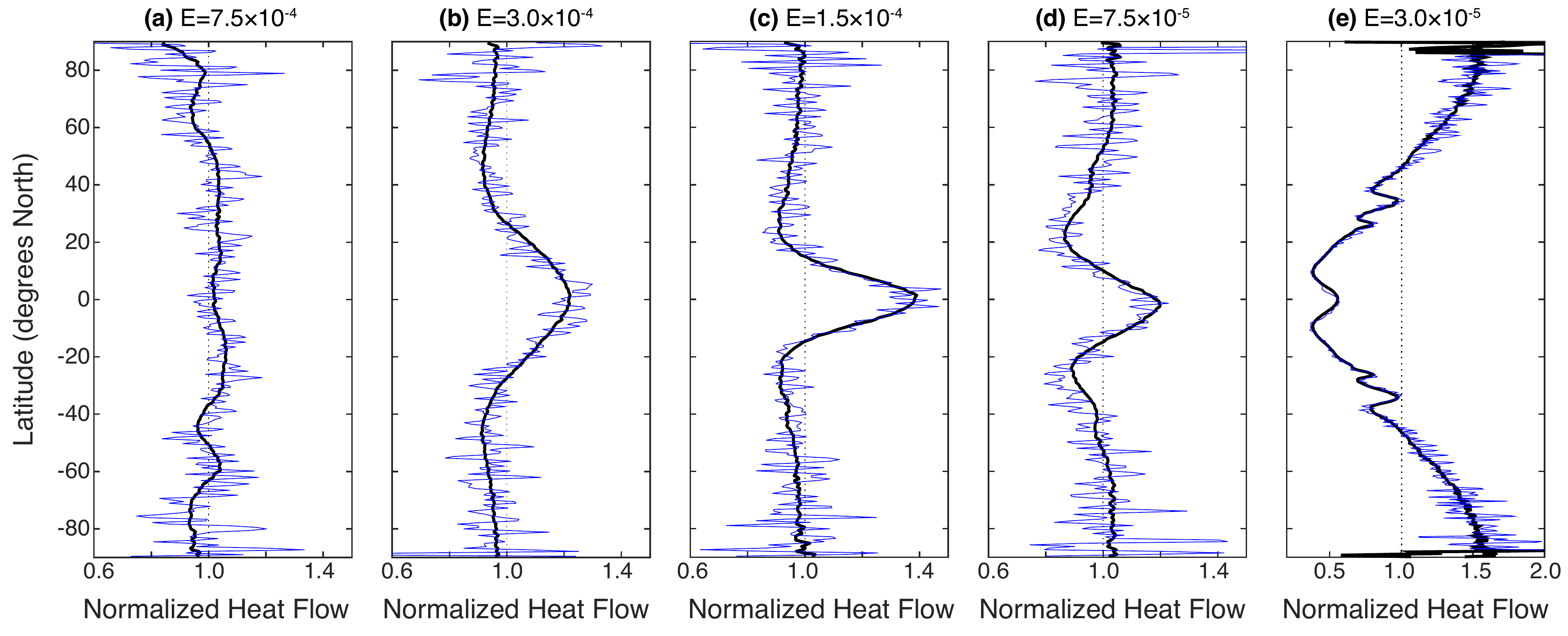}
\caption{Heat flux along the outer boundary normalized by the mean value for each model in the Ekman number suite. Black lines denote the average over all longitudes and at least 0.02 viscous diffusion times; blue lines show azimuthal averages at snapshots in time.}
\label{fig:HF}
\end{figure}

For Enceladus, I predict the zonal flows to be characterized by multiple jets that alternate in direction (Fig.~\ref{fig:Vel}A d-e). Converting model velocities to dimensional units $U = \Omega D Ro$, I expect peak zonal speeds of  nearly a m/s depending on the ocean thickness assumed. Meridional circulations are predicted to either be strongly aligned with the rotation axis with speeds up to a few mm/s (Fig.~\ref{fig:Vel}B e) or be concentrated in a low latitude upwelling with speeds up to a few cm/s (Fig.~\ref{fig:Vel}B d). As a result, heat flow along the ice-ocean interface has distinct peaks at either the poles (Figs.~\ref{fig:Vel}C e, \ref{fig:HF}e) or at the equator and the poles secondarily (Figs.~\ref{fig:Vel}C d, \ref{fig:HF}d).

Measurements of Enceladus' shape, gravitational field, and librational motions show that the ice shell is thin below the south pole and thick at the equator, with an intermediate thickness at the north pole \citep[e.g.,][]{CadekEA16,BeutheEA16}. Inverting these measurements to infer the oceanic heat flux along the ice-ocean interface, \citet[][]{CadekEA19} find peak flux near the poles with a minima at the equator. This pattern implies upwelling of warm water at the poles and downwelling of cool water at low latitudes, which may be caused by ocean convection (Fig.~\ref{fig:Vel}e) and/or be a consequence of the pattern of tidal heating in the mantle \citep{ChobletEA17}. Considering the former, the $E = 3.0 \times 10^{-5}$ model is appropriate if a low internal heat flux is assumed \citep[in contrast to][]{ChobletEA17} or if the thermal expansion coefficient in our calculations is overestimated since $\alpha$ trends towards zero with decreasing salinity and becomes negative for freshwater \citep[e.g.,][see also Table~\ref{tab:physparam}]{NayarEA16,Feistel10}, which both effectively reduce the Rayleigh number and make rotational effects more important.

Europa's ocean is predicted to have three zonal jets with retrograde equatorial flow that can reach m/s speeds (Fig.~\ref{fig:Vel}A b-c) or multiple zonal jets with retrograde equatorial flow and reduced speeds (Fig.~\ref{fig:Vel}A d) depending on the scaling law assumed. All Europa-relevant models, however, have an equatorial upwelling of warm water with peak speeds of roughly a few cm/s and enhanced heat transfer at low latitudes (Figs.~\ref{fig:Vel}B-C b-d, \ref{fig:HF}b-d).

The surface of Europa is riddled with geologic features indicating recent activity and the potential for ocean-derived materials \citep[e.g.,][]{FigueredoGreeley03,FischerEA15}. Chaos terrains, for example, appear to be located preferentially at low latitudes with a secondary prevalence near the poles \citep{LeonardEA18}, and formation models suggest that they may be associated with upwelling diapirs \citep[e.g.,][]{SotinEA02,CollinsNimmo09,SchmidtEA11} and marine ice accretion \citep[][]{SoderlundEA14}. No large gradients in ice shell thickness have been detected \citep{NimmoEA07}, suggesting an efficient ice pump \citep[c.f.][]{Nimmo04}. Given our robust model predictions of high oceanic heat flux at low latitudes with relatively low flux at mid-latitudes, our new calculations continue to support the thermocompositional diapirism  hypothesis. 

Given the similarities in regime predictions for Titan and Ganymede, they are considered together here. Assuming the $Ra^{Ga16}_{UNR}$ scaling and upper $Ra$ estimates, the oceans would behave akin to a non-rotating system with no coherent heat transfer patterns (Figs.~\ref{fig:Vel} C a, \ref{fig:HF}a). If the lower $Ra$ estimates are instead assumed, these satellites are predicted to have three-jet zonal flows with peak speeds up to a few m/s (depending on ocean thickness), Hadley-like circulation cells with peak speeds up to tens of cm/s (depending on ocean thickness), and maximum heat transfer near the equator (Figs.~\ref{fig:Vel} A-C b, \ref{fig:HF}b). Alternatively, for the $Ra^{RoC=1}_{UNR}$ scaling, these satellites are predicted to behave similarly to Enceladus and Europa as discussed above (Figs.~\ref{fig:Vel} A-C c-d, \ref{fig:HF}c-d), except with respect to dimensionalized flow speeds that could be considerably faster due to the larger ocean thicknesses.

Looking to Titan, the satellite's surface topography shows polar depressions compared to relatively elevated low latitudes \citep[e.g.,][]{DuranteEA19} that are likely explained by ice shell thickness variations \citep{NimmoBills10,HemingwayEA13,LefevreEA14} or ice shell density variations \citep{ChoukrounSotin12}. As for Enceladus, geophysical measurements by {\it Cassini} have been used to infer the oceanic heat flux along the ice-ocean interface \citep{KvorkaEA18}. The pattern is spatially complex, but simplifies to peaks near the poles when only axisymmetric components are considered. In contrast, the ocean convection models predicted to be relevant for Titan have either no coherent heat flux pattern (Fig.~\ref{fig:HF}a), peak flux and enhanced melting near the equator (Fig.~\ref{fig:HF}b-c), or peak flux near both the equator and the poles (Fig.~\ref{fig:HF}d), none of which are consistent with the observed long-wavelength topography assuming Airy isostasy. If ocean dynamics alone are responsible, this difference implies that either (1) the melted equatorial region in the intermediate scenario was infilled with less dense marine ice to form the equatorial bulge through Pratt isostasy or (2) the ocean has a stably-stratified salinity gradient that reduces the effective buoyancy forcing of the ocean ($Ra$) such that rotational effects become sufficient to maximize heat flow and melting at the poles (Fig.~\ref{fig:HF}e). %Alternatively, heat flow variations along the seafloor may be important.

Observational constraints for Ganymede are much more limited. The satellite's ancient grooved terrains indicate a likely period of geologic activity in its early history \citep{Lucchita80} and detection of hydrated salts suggests a subsurface briny layer of fluid \citep{McCordEA01}, but no clear patterns are present. Mass anomalies were measured in the northern hemisphere during a single Galileo flyby \citep{PalgutaEA06}, but the sparsity of data prohibit both characterization on a global scale and unique determination of their depth of origin. Consequently, there is no clear link at present between observations and the underlying ocean dynamics. 

\section{Discussion}

Our results are broadly consistent with the literature. Moreover, by comparing our numerical models against those with different input parameters, we are able to assess their sensitivity to these aspects. For example, the satellite oceans are predicted to have geometries characterized by $\chi$ values ranging from $0.74$ to $0.99$ (Table~\ref{tab:physparam}), compared to our models with fixed $\chi=0.90$. \cite{GastineEA13} found that anelastic columnar convection in thicker spherical shell geometries ($\chi=0.6$) is also characterized by a prograde equatorial jet with multiple, small-scale meridional circulations aligned with the rotation axis, which transitions to a regime with a retrograde equatorial jet and Hadley cell-like meridional circulations and ultimately a weakening of zonal flow speeds as the influence of rotation is decreased. Similarly, \cite{AurnouEA08} showed that heat transfer is inhibited at low latitudes and generally increases towards the poles for columnar convection in spherical shells with $\chi=[0.85, 0.9]$; this result is different from \cite{MiquelEA18}, who obtained peak equatorial heat transfer in their asymptotic models of rapidly rotating convection near onset, where the polar regions are subcritical \citep[e.g.,][]{DormyEA04}. Conversely, \cite{BrunPalacios09} and \cite{SoderlundEA13} showed that the equatorial heat transfer enhancement for less vertically stiff convection is robust for thicker shells ($\chi \leq 0.75$) and different thermal boundary conditions. Simulations with thinner spherical shells are computationally demanding and uncommon \citep[c.f.][]{DeRosaEA02}. Furthermore, studies with a thin layer of stable stratification below the outer boundary, which may be expected in regions where thermal expansivity is negative \citep[][see also Table~\ref{tab:physparam}]{MeloshEA04}, generally show similar trends \citep[e.g.,][]{HeimpelEA15}. Thus, these studies suggest that our results are not strongly sensitive to variations in ocean thickness or fluid properties with depth. Large spatial variations in ice shell thickness may enhance mechanically driven flows though \citep[e.g.,][]{LemasquerierEA17}, which are not considered here.

The effects of different boundary conditions should also be considered. For example, we assumed stress-free mechanical boundaries in order to reduce the effects of viscosity (e.g., artificially large Ekman boundary layers) due to the large $E$ values of our models compared to the satellites \citep[Table~\ref{tab:physparam};][]{KuangBloxham97}. In models with no-slip boundary conditions, inertial effects tend to be reduced substantially and strong zonal flows can be inhibited, which can disrupt convection \citep[e.g.,][]{AurnouHeimpel04,JonesTOG_2015}. For sufficiently driven and rapidly rotating convection, however, no-slip boundaries do not necessarily have this inhibiting effect \citep{MannevilleOlson96,AubertEA01}. Uniform fixed temperature boundary conditions were assumed because they enable a broader comparison with the literature and across the satellites, although fixed heat flux boundary conditions may be more appropriate along the seafloor. At moderate parameters, fixed flux conditions tend to promote larger convective scales \citep{SakurabaRoberts09,HoriEA12} and spatial variations along the boundary can influence the flow and efficiency of heat transfer, especially near the interface \citep[e.g.,][for anomalies along the outer boundary]{DietrichEA16,MoundDavies17}. At extreme (i.e. realistic) parameters, however, the solutions for both thermal conditions appear to converge for rapidly rotating convection \citep{CalkinsEA15_BC} as well as Rayleigh-B\'enard convection \citep{JohnstonDoering09}.  This convergence implies that convective-scale spatial variations in boundary heat flow have a secondary influence on the interior convection \citep{CalkinsEA15_BC}. While significant effects may occur if the spatial scale of the thermal anomaly is comparable to the vertical scale of convection \citep[e.g.,][]{DaviesEA09}, it is unclear whether these effects will persist across the entire fluid depth \citep{DaviesMound19}.

Future numerical work should (1) strive for more realistic Ekman and Rayleigh numbers, (2) tackle the effect of boundary conditions, especially the Stephan-type condition at the top boundary due to melting/freezing of water along the interface and fixed heat flux along the bottom boundary, (3) consider both temperature and salinity buoyancy sources \citep[e.g.,][]{VanceBrown05,Jansen16}, and (4) couple convectively and mechanically driven flows \citep[e.g.,][]{LeBarsEA15}.

Future missions to the outer solar system may be able to better constrain the ocean flows and test the predictions of our calculations and convection models. Looking specifically to the Jovian system, the {\it Europa Clipper} and {\it JUICE} missions will determine the ocean thickness and salinity and may be able to place constraints on spatial variations of ice shell thickness \citep[e.g.,][]{PhillipsPappalardo14,GrassetEA13}. Ice penetrating radar will provide information on ice shell thermophysical structure and constrain ice-ocean exchange processes \citep[e.g.,][]{KalousovaEA17}, while magnetometer measurements may allow probing of ocean currents through their induction of magnetic fields \citep[e.g.,][]{Tyler11}.

\acknowledgments
I thank Jonathan Aurnou, Baptiste Journaux, and Steve Vance for their helpful comments as well as Gabriel Tobie and Christophe Sotin for their constructive reviews. This work was supported by NASA Grant NNX14AR28G. Computational resources were provided by the NASA High-End Computing (HEC) Program through the NASA Advanced Supercomputing (NAS) Division at Ames Research Center. The MagIC code is publicly available at https://magic-sph.github.io/contents.html. All data is provided within the publication pages.

%% ------------------------------------------------------------------------ %%
%% References and Citations

%%%%%%%%%%%%%%%%%%%%%%%%%%%%%%%%%%%%%%%%%%%%%%%
% BibTeX is preferred:
%
%\bibliography{UCLA_UTIG.bib}
\bibliography{biblio_combined.bib}

\begin{thebibliography}{}

\bibitem [\protect \citeauthoryear {%
Abramson%
, Brown%
\BCBL {}\ \BBA {} Slutsky%
}{%
Abramson%
\ \protect \BOthers {.}}{%
{\protect \APACyear {2001}}%
}]{%
AbramsonEA01}
\APACinsertmetastar {%
AbramsonEA01}%
\begin{APACrefauthors}%
Abramson, E\BPBI H.%
, Brown, J\BPBI M.%
\BCBL {}\ \BBA {} Slutsky, L\BPBI J.%
\end{APACrefauthors}%
\unskip\
\newblock
\APACrefYearMonthDay{2001}{}{}.
\newblock
{\BBOQ}\APACrefatitle {{The thermal diffusivity of water at high pressures and
  temperatures}} {{The thermal diffusivity of water at high pressures and
  temperatures}}.{\BBCQ}
\newblock
\APACjournalVolNumPages{J. Chem. Phys.}{115}{}{10461--10463}.
\PrintBackRefs{\CurrentBib}

\bibitem [\protect \citeauthoryear {%
Aubert%
, Brito%
, Nataf%
, Cardin%
\BCBL {}\ \BBA {} Masson%
}{%
Aubert%
\ \protect \BOthers {.}}{%
{\protect \APACyear {2001}}%
}]{%
AubertEA01}
\APACinsertmetastar {%
AubertEA01}%
\begin{APACrefauthors}%
Aubert, J.%
, Brito, D.%
, Nataf, H\BPBI C.%
, Cardin, P.%
\BCBL {}\ \BBA {} Masson, J\BPBI P.%
\end{APACrefauthors}%
\unskip\
\newblock
\APACrefYearMonthDay{2001}{}{}.
\newblock
{\BBOQ}\APACrefatitle {A systematic experimental study of rapidly rotating
  spherical convection in water and liquid gallium} {A systematic experimental
  study of rapidly rotating spherical convection in water and liquid
  gallium}.{\BBCQ}
\newblock
\APACjournalVolNumPages{Phys. Earth Planet. Int.}{128}{1-4}{51--74}.
\PrintBackRefs{\CurrentBib}

\bibitem [\protect \citeauthoryear {%
Aurnou%
\ \protect \BOthers {.}}{%
Aurnou%
\ \protect \BOthers {.}}{%
{\protect \APACyear {2015}}%
}]{%
AurnouEA15}
\APACinsertmetastar {%
AurnouEA15}%
\begin{APACrefauthors}%
Aurnou, J\BPBI M.%
, Calkins, M\BPBI A.%
, Cheng, J\BPBI S.%
, Julien, K.%
, King, E\BPBI M.%
, Nieves, D.%
\BDBL {}Stellmach, S.%
\end{APACrefauthors}%
\unskip\
\newblock
\APACrefYearMonthDay{2015}{}{}.
\newblock
{\BBOQ}\APACrefatitle {Rotating convective turbulence in Earth and planetary
  cores} {Rotating convective turbulence in earth and planetary cores}.{\BBCQ}
\newblock
\APACjournalVolNumPages{Phys. Earth Planet. Int.}{246}{}{52--71}.
\PrintBackRefs{\CurrentBib}

\bibitem [\protect \citeauthoryear {%
Aurnou%
\ \BBA {} Heimpel%
}{%
Aurnou%
\ \BBA {} Heimpel%
}{%
{\protect \APACyear {2004}}%
}]{%
AurnouHeimpel04}
\APACinsertmetastar {%
AurnouHeimpel04}%
\begin{APACrefauthors}%
Aurnou, J\BPBI M.%
\BCBT {}\ \BBA {} Heimpel, M\BPBI H.%
\end{APACrefauthors}%
\unskip\
\newblock
\APACrefYearMonthDay{2004}{}{}.
\newblock
{\BBOQ}\APACrefatitle {Zonal jets in rotating convection with mixed mechanical
  boundary conditions} {Zonal jets in rotating convection with mixed mechanical
  boundary conditions}.{\BBCQ}
\newblock
\APACjournalVolNumPages{Icarus}{169}{}{492--498}.
\PrintBackRefs{\CurrentBib}

\bibitem [\protect \citeauthoryear {%
Aurnou%
, Heimpel%
, Allen%
, King%
\BCBL {}\ \BBA {} Wicht%
}{%
Aurnou%
\ \protect \BOthers {.}}{%
{\protect \APACyear {2008}}%
}]{%
AurnouEA08}
\APACinsertmetastar {%
AurnouEA08}%
\begin{APACrefauthors}%
Aurnou, J\BPBI M.%
, Heimpel, M\BPBI H.%
, Allen, L.%
, King, E\BPBI M.%
\BCBL {}\ \BBA {} Wicht, J.%
\end{APACrefauthors}%
\unskip\
\newblock
\APACrefYearMonthDay{2008}{}{}.
\newblock
{\BBOQ}\APACrefatitle {{Convective heat transfer and the pattern of thermal
  emission on the gas giants}} {{Convective heat transfer and the pattern of
  thermal emission on the gas giants}}.{\BBCQ}
\newblock
\APACjournalVolNumPages{Geophys. J. Int.}{173}{}{793--801}.
\PrintBackRefs{\CurrentBib}

\bibitem [\protect \citeauthoryear {%
Aurnou%
, Heimpel%
\BCBL {}\ \BBA {} Wicht%
}{%
Aurnou%
\ \protect \BOthers {.}}{%
{\protect \APACyear {2007}}%
}]{%
AurnouEA07}
\APACinsertmetastar {%
AurnouEA07}%
\begin{APACrefauthors}%
Aurnou, J\BPBI M.%
, Heimpel, M\BPBI H.%
\BCBL {}\ \BBA {} Wicht, J.%
\end{APACrefauthors}%
\unskip\
\newblock
\APACrefYearMonthDay{2007}{}{}.
\newblock
{\BBOQ}\APACrefatitle {{The effects of vigorous mixing in a convective model of
  zonal flow on the Ice Giants}} {{The effects of vigorous mixing in a
  convective model of zonal flow on the Ice Giants}}.{\BBCQ}
\newblock
\APACjournalVolNumPages{Icarus}{190}{}{110--126}.
\PrintBackRefs{\CurrentBib}

\bibitem [\protect \citeauthoryear {%
Aurnou%
\ \BBA {} Olson%
}{%
Aurnou%
\ \BBA {} Olson%
}{%
{\protect \APACyear {2001}}%
}]{%
AurnouOlson01}
\APACinsertmetastar {%
AurnouOlson01}%
\begin{APACrefauthors}%
Aurnou, J\BPBI M.%
\BCBT {}\ \BBA {} Olson, P\BPBI L.%
\end{APACrefauthors}%
\unskip\
\newblock
\APACrefYearMonthDay{2001}{}{}.
\newblock
{\BBOQ}\APACrefatitle {Strong zonal winds from thermal convection in a rotating
  spherical shell} {Strong zonal winds from thermal convection in a rotating
  spherical shell}.{\BBCQ}
\newblock
\APACjournalVolNumPages{Geophys. Res. Lett.}{28}{13}{2557--2559}.
\PrintBackRefs{\CurrentBib}

\bibitem [\protect \citeauthoryear {%
Baland%
, Tobie%
, Lefevre%
\BCBL {}\ \BBA {} Van~Hoolst%
}{%
Baland%
\ \protect \BOthers {.}}{%
{\protect \APACyear {2014}}%
}]{%
BalandEA14}
\APACinsertmetastar {%
BalandEA14}%
\begin{APACrefauthors}%
Baland, R\BHBI M.%
, Tobie, G.%
, Lefevre, A.%
\BCBL {}\ \BBA {} Van~Hoolst, T.%
\end{APACrefauthors}%
\unskip\
\newblock
\APACrefYearMonthDay{2014}{}{}.
\newblock
{\BBOQ}\APACrefatitle {{Titan's internal structure inferred from its gravity
  field, shape, and rotation state}} {{Titan's internal structure inferred from
  its gravity field, shape, and rotation state}}.{\BBCQ}
\newblock
\APACjournalVolNumPages{Icarus}{237}{}{29--41}.
\PrintBackRefs{\CurrentBib}

\bibitem [\protect \citeauthoryear {%
Beuthe%
, Rivoldini%
\BCBL {}\ \BBA {} Trinh%
}{%
Beuthe%
\ \protect \BOthers {.}}{%
{\protect \APACyear {2016}}%
}]{%
BeutheEA16}
\APACinsertmetastar {%
BeutheEA16}%
\begin{APACrefauthors}%
Beuthe, M.%
, Rivoldini, A.%
\BCBL {}\ \BBA {} Trinh, A.%
\end{APACrefauthors}%
\unskip\
\newblock
\APACrefYearMonthDay{2016}{}{}.
\newblock
{\BBOQ}\APACrefatitle {{Enceladus's and Dione's floating ice shells supported
  by minimum stress isostasy}} {{Enceladus's and Dione's floating ice shells
  supported by minimum stress isostasy}}.{\BBCQ}
\newblock
\APACjournalVolNumPages{Geophys. Res. Lett.}{43}{19}{}.
\PrintBackRefs{\CurrentBib}

\bibitem [\protect \citeauthoryear {%
Bland%
, Showman%
\BCBL {}\ \BBA {} Tobie%
}{%
Bland%
\ \protect \BOthers {.}}{%
{\protect \APACyear {2009}}%
}]{%
BlandEA09}
\APACinsertmetastar {%
BlandEA09}%
\begin{APACrefauthors}%
Bland, M\BPBI T.%
, Showman, A\BPBI P.%
\BCBL {}\ \BBA {} Tobie, G.%
\end{APACrefauthors}%
\unskip\
\newblock
\APACrefYearMonthDay{2009}{}{}.
\newblock
{\BBOQ}\APACrefatitle {{The orbital--thermal evolution and global expansion of
  Ganymede}} {{The orbital--thermal evolution and global expansion of
  Ganymede}}.{\BBCQ}
\newblock
\APACjournalVolNumPages{Icarus}{200}{1}{207--221}.
\PrintBackRefs{\CurrentBib}

\bibitem [\protect \citeauthoryear {%
Brun%
\ \BBA {} Palacios%
}{%
Brun%
\ \BBA {} Palacios%
}{%
{\protect \APACyear {2009}}%
}]{%
BrunPalacios09}
\APACinsertmetastar {%
BrunPalacios09}%
\begin{APACrefauthors}%
Brun, A\BPBI S.%
\BCBT {}\ \BBA {} Palacios, A.%
\end{APACrefauthors}%
\unskip\
\newblock
\APACrefYearMonthDay{2009}{}{}.
\newblock
{\BBOQ}\APACrefatitle {{Numerical simulations of a rotating red giant star.
  {I}. {T}hree-dimensional models of turbulent convection and associated mean
  flows}} {{Numerical simulations of a rotating red giant star. {I}.
  {T}hree-dimensional models of turbulent convection and associated mean
  flows}}.{\BBCQ}
\newblock
\APACjournalVolNumPages{Astrophys. J.}{702}{}{1078--1097}.
\PrintBackRefs{\CurrentBib}

\bibitem [\protect \citeauthoryear {%
{\v{C}}adek%
\ \protect \BOthers {.}}{%
{\v{C}}adek%
\ \protect \BOthers {.}}{%
{\protect \APACyear {2019}}%
}]{%
CadekEA19}
\APACinsertmetastar {%
CadekEA19}%
\begin{APACrefauthors}%
{\v{C}}adek, O.%
, Sou{\v{c}}ek, O.%
, B{\v{e}}hounkov{\'a}, M.%
, Choblet, G.%
, Tobie, G.%
\BCBL {}\ \BBA {} Hron, J.%
\end{APACrefauthors}%
\unskip\
\newblock
\APACrefYearMonthDay{2019}{}{}.
\newblock
{\BBOQ}\APACrefatitle {{Long-term stability of Enceladus' uneven ice shell}}
  {{Long-term stability of Enceladus' uneven ice shell}}.{\BBCQ}
\newblock
\APACjournalVolNumPages{Icarus}{319}{}{476--484}.
\PrintBackRefs{\CurrentBib}

\bibitem [\protect \citeauthoryear {%
{\v{C}}adek%
\ \protect \BOthers {.}}{%
{\v{C}}adek%
\ \protect \BOthers {.}}{%
{\protect \APACyear {2016}}%
}]{%
CadekEA16}
\APACinsertmetastar {%
CadekEA16}%
\begin{APACrefauthors}%
{\v{C}}adek, O.%
, Tobie, G.%
, Van~Hoolst, T.%
, Mass{\'e}, M.%
, Choblet, G.%
, Lef{\`e}vre, A.%
\BDBL {}others%
\end{APACrefauthors}%
\unskip\
\newblock
\APACrefYearMonthDay{2016}{}{}.
\newblock
{\BBOQ}\APACrefatitle {{Enceladus's internal ocean and ice shell constrained
  from Cassini gravity, shape, and libration data}} {{Enceladus's internal
  ocean and ice shell constrained from Cassini gravity, shape, and libration
  data}}.{\BBCQ}
\newblock
\APACjournalVolNumPages{Geophys. Res. Lett.}{43}{11}{5653--5660}.
\PrintBackRefs{\CurrentBib}

\bibitem [\protect \citeauthoryear {%
Calkins%
\ \protect \BOthers {.}}{%
Calkins%
\ \protect \BOthers {.}}{%
{\protect \APACyear {2015}}%
}]{%
CalkinsEA15_BC}
\APACinsertmetastar {%
CalkinsEA15_BC}%
\begin{APACrefauthors}%
Calkins, M\BPBI A.%
, Hale, K.%
, Julien, K.%
, Nieves, D.%
, Driggs, D.%
\BCBL {}\ \BBA {} Marti, P.%
\end{APACrefauthors}%
\unskip\
\newblock
\APACrefYearMonthDay{2015}{}{}.
\newblock
{\BBOQ}\APACrefatitle {The asymptotic equivalence of fixed heat flux and fixed
  temperature thermal boundary conditions for rapidly rotating convection} {The
  asymptotic equivalence of fixed heat flux and fixed temperature thermal
  boundary conditions for rapidly rotating convection}.{\BBCQ}
\newblock
\APACjournalVolNumPages{J. Fluid Mech.}{784}{}{R2}.
\PrintBackRefs{\CurrentBib}

\bibitem [\protect \citeauthoryear {%
{Chandrasekhar}%
}{%
{Chandrasekhar}%
}{%
{\protect \APACyear {1961}}%
}]{%
Chandrasekhar61}
\APACinsertmetastar {%
Chandrasekhar61}%
\begin{APACrefauthors}%
{Chandrasekhar}, S.%
\end{APACrefauthors}%
\unskip\
\newblock
\APACrefYear{1961}.
\newblock
\APACrefbtitle {{Hydrodynamic and hydromagnetic stability}} {{Hydrodynamic and
  hydromagnetic stability}}.
\newblock
\APACaddressPublisher{}{Oxford: Clarendon}.
\PrintBackRefs{\CurrentBib}

\bibitem [\protect \citeauthoryear {%
Cheng%
, Aurnou%
, Julien%
\BCBL {}\ \BBA {} Kunnen%
}{%
Cheng%
\ \protect \BOthers {.}}{%
{\protect \APACyear {2018}}%
}]{%
ChengEA18}
\APACinsertmetastar {%
ChengEA18}%
\begin{APACrefauthors}%
Cheng, J\BPBI S.%
, Aurnou, J\BPBI M.%
, Julien, K.%
\BCBL {}\ \BBA {} Kunnen, R\BPBI P\BPBI J.%
\end{APACrefauthors}%
\unskip\
\newblock
\APACrefYearMonthDay{2018}{}{}.
\newblock
{\BBOQ}\APACrefatitle {A heuristic framework for next-generation models of
  geostrophic convective turbulence} {A heuristic framework for next-generation
  models of geostrophic convective turbulence}.{\BBCQ}
\newblock
\APACjournalVolNumPages{Geophys. Astrophys. Fluid Dyn.}{112}{4}{277--300}.
\PrintBackRefs{\CurrentBib}

\bibitem [\protect \citeauthoryear {%
Choblet%
\ \protect \BOthers {.}}{%
Choblet%
\ \protect \BOthers {.}}{%
{\protect \APACyear {2017}}%
}]{%
ChobletEA17}
\APACinsertmetastar {%
ChobletEA17}%
\begin{APACrefauthors}%
Choblet, G.%
, Tobie, G.%
, Sotin, C.%
, B{\v{e}}hounkov{\'{a}}, M.%
, {\v{C}}adek, O.%
, Postberg, F.%
\BCBL {}\ \BBA {} Sou{\v{c}}ek, O.%
\end{APACrefauthors}%
\unskip\
\newblock
\APACrefYearMonthDay{2017}{}{}.
\newblock
{\BBOQ}\APACrefatitle {{Powering prolonged hydrothermal activity inside
  Enceladus}} {{Powering prolonged hydrothermal activity inside
  Enceladus}}.{\BBCQ}
\newblock
\APACjournalVolNumPages{Nature Astron.}{1}{12}{841}.
\PrintBackRefs{\CurrentBib}

\bibitem [\protect \citeauthoryear {%
Choukroun%
\ \BBA {} Sotin%
}{%
Choukroun%
\ \BBA {} Sotin%
}{%
{\protect \APACyear {2012}}%
}]{%
ChoukrounSotin12}
\APACinsertmetastar {%
ChoukrounSotin12}%
\begin{APACrefauthors}%
Choukroun, M.%
\BCBT {}\ \BBA {} Sotin, C.%
\end{APACrefauthors}%
\unskip\
\newblock
\APACrefYearMonthDay{2012}{}{}.
\newblock
{\BBOQ}\APACrefatitle {{Is Titan's shape caused by its meteorology and carbon
  cycle?}} {{Is Titan's shape caused by its meteorology and carbon
  cycle?}}{\BBCQ}
\newblock
\APACjournalVolNumPages{Geophys. Res. Lett.}{39}{4}{}.
\PrintBackRefs{\CurrentBib}

\bibitem [\protect \citeauthoryear {%
Christensen%
}{%
Christensen%
}{%
{\protect \APACyear {2001}}%
}]{%
Christensen01}
\APACinsertmetastar {%
Christensen01}%
\begin{APACrefauthors}%
Christensen, U\BPBI R.%
\end{APACrefauthors}%
\unskip\
\newblock
\APACrefYearMonthDay{2001}{}{}.
\newblock
{\BBOQ}\APACrefatitle {{Zonal flow driven by deep convection on the major
  planets}} {{Zonal flow driven by deep convection on the major
  planets}}.{\BBCQ}
\newblock
\APACjournalVolNumPages{Geophys. Res. Lett.}{28}{}{2553--2556}.
\PrintBackRefs{\CurrentBib}

\bibitem [\protect \citeauthoryear {%
Collins%
\ \BBA {} Nimmo%
}{%
Collins%
\ \BBA {} Nimmo%
}{%
{\protect \APACyear {2009}}%
}]{%
CollinsNimmo09}
\APACinsertmetastar {%
CollinsNimmo09}%
\begin{APACrefauthors}%
Collins, G.%
\BCBT {}\ \BBA {} Nimmo, F.%
\end{APACrefauthors}%
\unskip\
\newblock
\APACrefYearMonthDay{2009}{}{}.
\newblock
{\BBOQ}\APACrefatitle {Chaotic terrain on {E}uropa} {Chaotic terrain on
  {E}uropa}.{\BBCQ}
\newblock
\BIn{} R\BPBI T.~Pappalardo, W\BPBI B.~McKinnon\BCBL {}\ \BBA {} K\BPBI
  K.~Khurana\ (\BEDS), \APACrefbtitle {{E}uropa} {{E}uropa}\ (\BPG~259-281).
\newblock
\APACaddressPublisher{}{Tucson: University of Arizona Press}.
\PrintBackRefs{\CurrentBib}

\bibitem [\protect \citeauthoryear {%
Daly%
, Rack%
\BCBL {}\ \BBA {} Zook%
}{%
Daly%
\ \protect \BOthers {.}}{%
{\protect \APACyear {2013}}%
}]{%
DalyEA13}
\APACinsertmetastar {%
DalyEA13}%
\begin{APACrefauthors}%
Daly, M.%
, Rack, F.%
\BCBL {}\ \BBA {} Zook, R.%
\end{APACrefauthors}%
\unskip\
\newblock
\APACrefYearMonthDay{2013}{}{}.
\newblock
{\BBOQ}\APACrefatitle {{Edwardsiella andrillae, a new species of sea anemone
  from Antarctic Ice}} {{Edwardsiella andrillae, a new species of sea anemone
  from Antarctic Ice}}.{\BBCQ}
\newblock
\APACjournalVolNumPages{PloS one}{8}{12}{e83476}.
\PrintBackRefs{\CurrentBib}

\bibitem [\protect \citeauthoryear {%
Davies%
, Gubbins%
\BCBL {}\ \BBA {} Jimack%
}{%
Davies%
\ \protect \BOthers {.}}{%
{\protect \APACyear {2009}}%
}]{%
DaviesEA09}
\APACinsertmetastar {%
DaviesEA09}%
\begin{APACrefauthors}%
Davies, C\BPBI J.%
, Gubbins, D.%
\BCBL {}\ \BBA {} Jimack, P\BPBI K.%
\end{APACrefauthors}%
\unskip\
\newblock
\APACrefYearMonthDay{2009}{}{}.
\newblock
{\BBOQ}\APACrefatitle {Convection in a rapidly rotating spherical shell with an
  imposed laterally varying thermal boundary condition} {Convection in a
  rapidly rotating spherical shell with an imposed laterally varying thermal
  boundary condition}.{\BBCQ}
\newblock
\APACjournalVolNumPages{J. Fluid Mech.}{641}{}{335--358}.
\PrintBackRefs{\CurrentBib}

\bibitem [\protect \citeauthoryear {%
Davies%
\ \BBA {} Mound%
}{%
Davies%
\ \BBA {} Mound%
}{%
{\protect \APACyear {2019}}%
}]{%
DaviesMound19}
\APACinsertmetastar {%
DaviesMound19}%
\begin{APACrefauthors}%
Davies, C\BPBI J.%
\BCBT {}\ \BBA {} Mound, J\BPBI E.%
\end{APACrefauthors}%
\unskip\
\newblock
\APACrefYearMonthDay{2019}{}{}.
\newblock
{\BBOQ}\APACrefatitle {Mantle-induced Temperature Anomalies Do Not Reach The
  Inner Core Boundary} {Mantle-induced temperature anomalies do not reach the
  inner core boundary}.{\BBCQ}
\newblock
\APACjournalVolNumPages{Geophys. J. Int.}{218}{}{2054--065}.
\PrintBackRefs{\CurrentBib}

\bibitem [\protect \citeauthoryear {%
DeRosa%
, Gilman%
\BCBL {}\ \BBA {} Toomre%
}{%
DeRosa%
\ \protect \BOthers {.}}{%
{\protect \APACyear {2002}}%
}]{%
DeRosaEA02}
\APACinsertmetastar {%
DeRosaEA02}%
\begin{APACrefauthors}%
DeRosa, M\BPBI L.%
, Gilman, P\BPBI A.%
\BCBL {}\ \BBA {} Toomre, J.%
\end{APACrefauthors}%
\unskip\
\newblock
\APACrefYearMonthDay{2002}{}{}.
\newblock
{\BBOQ}\APACrefatitle {Solar multiscale convection and rotation gradients
  studied in shallow spherical shells} {Solar multiscale convection and
  rotation gradients studied in shallow spherical shells}.{\BBCQ}
\newblock
\APACjournalVolNumPages{Astrophys. J.}{581}{2}{1356}.
\PrintBackRefs{\CurrentBib}

\bibitem [\protect \citeauthoryear {%
Dietrich%
, Hori%
\BCBL {}\ \BBA {} Wicht%
}{%
Dietrich%
\ \protect \BOthers {.}}{%
{\protect \APACyear {2016}}%
}]{%
DietrichEA16}
\APACinsertmetastar {%
DietrichEA16}%
\begin{APACrefauthors}%
Dietrich, W.%
, Hori, K.%
\BCBL {}\ \BBA {} Wicht, J.%
\end{APACrefauthors}%
\unskip\
\newblock
\APACrefYearMonthDay{2016}{}{}.
\newblock
{\BBOQ}\APACrefatitle {Core flows and heat transfer induced by inhomogeneous
  cooling with sub-and supercritical convection} {Core flows and heat transfer
  induced by inhomogeneous cooling with sub-and supercritical
  convection}.{\BBCQ}
\newblock
\APACjournalVolNumPages{Phys. Earth Planet. Int.}{251}{}{36--51}.
\PrintBackRefs{\CurrentBib}

\bibitem [\protect \citeauthoryear {%
Dombard%
\ \BBA {} McKinnon%
}{%
Dombard%
\ \BBA {} McKinnon%
}{%
{\protect \APACyear {2001}}%
}]{%
DombardMcKinnon01}
\APACinsertmetastar {%
DombardMcKinnon01}%
\begin{APACrefauthors}%
Dombard, A\BPBI J.%
\BCBT {}\ \BBA {} McKinnon, W\BPBI B.%
\end{APACrefauthors}%
\unskip\
\newblock
\APACrefYearMonthDay{2001}{}{}.
\newblock
{\BBOQ}\APACrefatitle {{Formation of grooved terrain on Ganymede: Extensional
  instability mediated by cold, superplastic creep}} {{Formation of grooved
  terrain on Ganymede: Extensional instability mediated by cold, superplastic
  creep}}.{\BBCQ}
\newblock
\APACjournalVolNumPages{Icarus}{154}{2}{321--336}.
\PrintBackRefs{\CurrentBib}

\bibitem [\protect \citeauthoryear {%
Dormy%
, Soward%
, Jones%
, Jault%
\BCBL {}\ \BBA {} Cardin%
}{%
Dormy%
\ \protect \BOthers {.}}{%
{\protect \APACyear {2004}}%
}]{%
DormyEA04}
\APACinsertmetastar {%
DormyEA04}%
\begin{APACrefauthors}%
Dormy, E.%
, Soward, A\BPBI M.%
, Jones, C\BPBI A.%
, Jault, D.%
\BCBL {}\ \BBA {} Cardin, P.%
\end{APACrefauthors}%
\unskip\
\newblock
\APACrefYearMonthDay{2004}{}{}.
\newblock
{\BBOQ}\APACrefatitle {{The onset of thermal convection in rotating spherical
  shells}} {{The onset of thermal convection in rotating spherical
  shells}}.{\BBCQ}
\newblock
\APACjournalVolNumPages{J. Fluid Mech.}{501}{}{43--70}.
\PrintBackRefs{\CurrentBib}

\bibitem [\protect \citeauthoryear {%
Durante%
, Hemingway%
, Racioppa%
, Iess%
\BCBL {}\ \BBA {} Stevenson%
}{%
Durante%
\ \protect \BOthers {.}}{%
{\protect \APACyear {2019}}%
}]{%
DuranteEA19}
\APACinsertmetastar {%
DuranteEA19}%
\begin{APACrefauthors}%
Durante, D.%
, Hemingway, D\BPBI J.%
, Racioppa, P.%
, Iess, L.%
\BCBL {}\ \BBA {} Stevenson, D\BPBI J.%
\end{APACrefauthors}%
\unskip\
\newblock
\APACrefYearMonthDay{2019}{}{}.
\newblock
{\BBOQ}\APACrefatitle {{Titan's gravity field and interior structure after
  Cassini}} {{Titan's gravity field and interior structure after
  Cassini}}.{\BBCQ}
\newblock
\APACjournalVolNumPages{Icarus}{326}{}{123--132}.
\PrintBackRefs{\CurrentBib}

\bibitem [\protect \citeauthoryear {%
Ecke%
\ \BBA {} Niemela%
}{%
Ecke%
\ \BBA {} Niemela%
}{%
{\protect \APACyear {2014}}%
}]{%
EckeNiemela14}
\APACinsertmetastar {%
EckeNiemela14}%
\begin{APACrefauthors}%
Ecke, R\BPBI E.%
\BCBT {}\ \BBA {} Niemela, J\BPBI J.%
\end{APACrefauthors}%
\unskip\
\newblock
\APACrefYearMonthDay{2014}{}{}.
\newblock
{\BBOQ}\APACrefatitle {{Heat transport in the geostrophic regime of rotating
  Rayleigh-B\'enard convection}} {{Heat transport in the geostrophic regime of
  rotating Rayleigh-B\'enard convection}}.{\BBCQ}
\newblock
\APACjournalVolNumPages{Phys. Rev. Lett.}{113}{}{114301}.
\PrintBackRefs{\CurrentBib}

\bibitem [\protect \citeauthoryear {%
Eicken%
, Oerter%
, Miller%
\BCBL {}\ \BBA {} Graf%
}{%
Eicken%
\ \protect \BOthers {.}}{%
{\protect \APACyear {1984}}%
}]{%
EickenEA94}
\APACinsertmetastar {%
EickenEA94}%
\begin{APACrefauthors}%
Eicken, H.%
, Oerter, H.%
, Miller, H.%
\BCBL {}\ \BBA {} Graf, W.%
\end{APACrefauthors}%
\unskip\
\newblock
\APACrefYearMonthDay{1984}{}{}.
\newblock
{\BBOQ}\APACrefatitle {{Textural characteristics and impurity content of
  meteoric and marine ice in the Ronne Ice Shelf, Antarctica}} {{Textural
  characteristics and impurity content of meteoric and marine ice in the Ronne
  Ice Shelf, Antarctica}}.{\BBCQ}
\newblock
\APACjournalVolNumPages{J. Glaciol.}{40}{135}{386--398}.
\PrintBackRefs{\CurrentBib}

\bibitem [\protect \citeauthoryear {%
Feistel%
}{%
Feistel%
}{%
{\protect \APACyear {2010}}%
}]{%
Feistel10}
\APACinsertmetastar {%
Feistel10}%
\begin{APACrefauthors}%
Feistel, R.%
\end{APACrefauthors}%
\unskip\
\newblock
\APACrefYearMonthDay{2010}{}{}.
\newblock
{\BBOQ}\APACrefatitle {Extended equation of state for seawater at elevated
  temperature and salinity} {Extended equation of state for seawater at
  elevated temperature and salinity}.{\BBCQ}
\newblock
\APACjournalVolNumPages{Desalination}{250}{}{14--18}.
\PrintBackRefs{\CurrentBib}

\bibitem [\protect \citeauthoryear {%
Figueredo%
\ \BBA {} Greeley%
}{%
Figueredo%
\ \BBA {} Greeley%
}{%
{\protect \APACyear {2003}}%
}]{%
FigueredoGreeley03}
\APACinsertmetastar {%
FigueredoGreeley03}%
\begin{APACrefauthors}%
Figueredo, P\BPBI H.%
\BCBT {}\ \BBA {} Greeley, R.%
\end{APACrefauthors}%
\unskip\
\newblock
\APACrefYearMonthDay{2003}{}{}.
\newblock
{\BBOQ}\APACrefatitle {{The Emerging Resurfacing History of Europa from
  Pole-to-Pole Geologic Mapping}} {{The Emerging Resurfacing History of Europa
  from Pole-to-Pole Geologic Mapping}}.{\BBCQ}
\newblock
\BIn{} {{Mackwell}, S. and {Stansbery}, E.}\ (\BED), \APACrefbtitle {{Lunar and
  Planetary Institute Science Conference Abstracts}} {{Lunar and Planetary
  Institute Science Conference Abstracts}}\ (\BVOL~34, \BPG~1017).
\PrintBackRefs{\CurrentBib}

\bibitem [\protect \citeauthoryear {%
Fischer%
, Brown%
\BCBL {}\ \BBA {} Hand%
}{%
Fischer%
\ \protect \BOthers {.}}{%
{\protect \APACyear {2015}}%
}]{%
FischerEA15}
\APACinsertmetastar {%
FischerEA15}%
\begin{APACrefauthors}%
Fischer, P\BPBI D.%
, Brown, M\BPBI E.%
\BCBL {}\ \BBA {} Hand, K\BPBI P.%
\end{APACrefauthors}%
\unskip\
\newblock
\APACrefYearMonthDay{2015}{}{}.
\newblock
{\BBOQ}\APACrefatitle {{Spatially resolved spectroscopy of Europa: The distinct
  spectrum of large-scale chaos}} {{Spatially resolved spectroscopy of Europa:
  The distinct spectrum of large-scale chaos}}.{\BBCQ}
\newblock
\APACjournalVolNumPages{Astron. J.}{150}{5}{164}.
\PrintBackRefs{\CurrentBib}

\bibitem [\protect \citeauthoryear {%
Gastine%
, Heimpel%
\BCBL {}\ \BBA {} Wicht%
}{%
Gastine%
\ \protect \BOthers {.}}{%
{\protect \APACyear {2014}}%
}]{%
GastineEA14}
\APACinsertmetastar {%
GastineEA14}%
\begin{APACrefauthors}%
Gastine, T.%
, Heimpel, M\BPBI H.%
\BCBL {}\ \BBA {} Wicht, J.%
\end{APACrefauthors}%
\unskip\
\newblock
\APACrefYearMonthDay{2014}{}{}.
\newblock
{\BBOQ}\APACrefatitle {Zonal flow scaling in rapidly-rotating compressible
  convection} {Zonal flow scaling in rapidly-rotating compressible
  convection}.{\BBCQ}
\newblock
\APACjournalVolNumPages{Phys. Earth Planet. Int.}{232}{}{36--50}.
\PrintBackRefs{\CurrentBib}

\bibitem [\protect \citeauthoryear {%
Gastine%
\ \BBA {} Wicht%
}{%
Gastine%
\ \BBA {} Wicht%
}{%
{\protect \APACyear {2012}}%
}]{%
GastineWicht12}
\APACinsertmetastar {%
GastineWicht12}%
\begin{APACrefauthors}%
Gastine, T.%
\BCBT {}\ \BBA {} Wicht, J.%
\end{APACrefauthors}%
\unskip\
\newblock
\APACrefYearMonthDay{2012}{}{}.
\newblock
{\BBOQ}\APACrefatitle {Effects of compressibility on driving zonal flow in gas
  giants} {Effects of compressibility on driving zonal flow in gas
  giants}.{\BBCQ}
\newblock
\APACjournalVolNumPages{Icarus}{219}{}{428--442}.
\PrintBackRefs{\CurrentBib}

\bibitem [\protect \citeauthoryear {%
Gastine%
, Wicht%
\BCBL {}\ \BBA {} Aubert%
}{%
Gastine%
\ \protect \BOthers {.}}{%
{\protect \APACyear {2016}}%
}]{%
GastineEA16}
\APACinsertmetastar {%
GastineEA16}%
\begin{APACrefauthors}%
Gastine, T.%
, Wicht, J.%
\BCBL {}\ \BBA {} Aubert, J.%
\end{APACrefauthors}%
\unskip\
\newblock
\APACrefYearMonthDay{2016}{}{}.
\newblock
{\BBOQ}\APACrefatitle {{Scaling regimes in spherical shell rotating
  convection}} {{Scaling regimes in spherical shell rotating
  convection}}.{\BBCQ}
\newblock
\APACjournalVolNumPages{J. Fluid Mech.}{808}{}{690--732}.
\PrintBackRefs{\CurrentBib}

\bibitem [\protect \citeauthoryear {%
Gastine%
, Wicht%
\BCBL {}\ \BBA {} Aurnou%
}{%
Gastine%
\ \protect \BOthers {.}}{%
{\protect \APACyear {2013}}%
}]{%
GastineEA13}
\APACinsertmetastar {%
GastineEA13}%
\begin{APACrefauthors}%
Gastine, T.%
, Wicht, J.%
\BCBL {}\ \BBA {} Aurnou, J\BPBI M.%
\end{APACrefauthors}%
\unskip\
\newblock
\APACrefYearMonthDay{2013}{}{}.
\newblock
{\BBOQ}\APACrefatitle {Zonal flow regimes in rotating anelastic spherical
  shells: An application to giant planets} {Zonal flow regimes in rotating
  anelastic spherical shells: An application to giant planets}.{\BBCQ}
\newblock
\APACjournalVolNumPages{Icarus}{225}{}{156--172}.
\PrintBackRefs{\CurrentBib}

\bibitem [\protect \citeauthoryear {%
Gastine%
, Wicht%
\BCBL {}\ \BBA {} Aurnou%
}{%
Gastine%
\ \protect \BOthers {.}}{%
{\protect \APACyear {2015}}%
}]{%
GastineEA15}
\APACinsertmetastar {%
GastineEA15}%
\begin{APACrefauthors}%
Gastine, T.%
, Wicht, J.%
\BCBL {}\ \BBA {} Aurnou, J\BPBI M.%
\end{APACrefauthors}%
\unskip\
\newblock
\APACrefYearMonthDay{2015}{}{}.
\newblock
{\BBOQ}\APACrefatitle {{Turbulent Rayleigh--B{\'e}nard convection in spherical
  shells}} {{Turbulent Rayleigh--B{\'e}nard convection in spherical
  shells}}.{\BBCQ}
\newblock
\APACjournalVolNumPages{J. Fluid Mech.}{778}{}{721--764}.
\PrintBackRefs{\CurrentBib}

\bibitem [\protect \citeauthoryear {%
Gilman%
}{%
Gilman%
}{%
{\protect \APACyear {1977}}%
}]{%
Gilman77}
\APACinsertmetastar {%
Gilman77}%
\begin{APACrefauthors}%
Gilman, P\BPBI A.%
\end{APACrefauthors}%
\unskip\
\newblock
\APACrefYearMonthDay{1977}{}{}.
\newblock
{\BBOQ}\APACrefatitle {{Nonlinear dynamics of {B}oussinesq convection in a deep
  rotating spherical shell -- {I}}} {{Nonlinear dynamics of {B}oussinesq
  convection in a deep rotating spherical shell -- {I}}}.{\BBCQ}
\newblock
\APACjournalVolNumPages{Geophys. Astrophys. Fluid Dyn.}{8}{}{93--135}.
\PrintBackRefs{\CurrentBib}

\bibitem [\protect \citeauthoryear {%
Gilman%
}{%
Gilman%
}{%
{\protect \APACyear {1978}}%
}]{%
Gilman78}
\APACinsertmetastar {%
Gilman78}%
\begin{APACrefauthors}%
Gilman, P\BPBI A.%
\end{APACrefauthors}%
\unskip\
\newblock
\APACrefYearMonthDay{1978}{}{}.
\newblock
{\BBOQ}\APACrefatitle {{Nonlinear dynamics of {B}oussinesq convection in a deep
  rotating spherical shell -- {II}}} {{Nonlinear dynamics of {B}oussinesq
  convection in a deep rotating spherical shell -- {II}}}.{\BBCQ}
\newblock
\APACjournalVolNumPages{Geophys. Astrophys. Fluid Dyn.}{11}{}{157--179}.
\PrintBackRefs{\CurrentBib}

\bibitem [\protect \citeauthoryear {%
Gissinger%
\ \BBA {} Petitdemange%
}{%
Gissinger%
\ \BBA {} Petitdemange%
}{%
{\protect \APACyear {2019}}%
}]{%
GissingerPetitdemange19}
\APACinsertmetastar {%
GissingerPetitdemange19}%
\begin{APACrefauthors}%
Gissinger, C.%
\BCBT {}\ \BBA {} Petitdemange, L.%
\end{APACrefauthors}%
\unskip\
\newblock
\APACrefYearMonthDay{2019}{}{}.
\newblock
{\BBOQ}\APACrefatitle {{A magnetically driven equatorial jet in Europa's
  ocean}} {{A magnetically driven equatorial jet in Europa's ocean}}.{\BBCQ}
\newblock
\APACjournalVolNumPages{Nature Astron.}{1}{}{}.
\PrintBackRefs{\CurrentBib}

\bibitem [\protect \citeauthoryear {%
Goodman%
}{%
Goodman%
}{%
{\protect \APACyear {2014}}%
}]{%
Goodman14}
\APACinsertmetastar {%
Goodman14}%
\begin{APACrefauthors}%
Goodman, J\BPBI C.%
\end{APACrefauthors}%
\unskip\
\newblock
\APACrefYearMonthDay{2014}{}{}.
\newblock
{\BBOQ}\APACrefatitle {Flow of an alien ocean} {Flow of an alien ocean}.{\BBCQ}
\newblock
\APACjournalVolNumPages{Nature Geosci.}{7}{}{8--9}.
\PrintBackRefs{\CurrentBib}

\bibitem [\protect \citeauthoryear {%
Grasset%
\ \protect \BOthers {.}}{%
Grasset%
\ \protect \BOthers {.}}{%
{\protect \APACyear {2013}}%
}]{%
GrassetEA13}
\APACinsertmetastar {%
GrassetEA13}%
\begin{APACrefauthors}%
Grasset, O.%
, Dougherty, M\BPBI K.%
, Coustenis, A.%
, Bunce, E\BPBI J.%
, Erd, C.%
, Titov, D.%
\BDBL {}others%
\end{APACrefauthors}%
\unskip\
\newblock
\APACrefYearMonthDay{2013}{}{}.
\newblock
{\BBOQ}\APACrefatitle {{JUpiter ICy moons Explorer (JUICE): An ESA mission to
  orbit Ganymede and to characterise the Jupiter system}} {{JUpiter ICy moons
  Explorer (JUICE): An ESA mission to orbit Ganymede and to characterise the
  Jupiter system}}.{\BBCQ}
\newblock
\APACjournalVolNumPages{Planet. Space Sci.}{78}{}{1--21}.
\PrintBackRefs{\CurrentBib}

\bibitem [\protect \citeauthoryear {%
Hartkorn%
\ \BBA {} Saur%
}{%
Hartkorn%
\ \BBA {} Saur%
}{%
{\protect \APACyear {2017}}%
}]{%
HartkornSaur17}
\APACinsertmetastar {%
HartkornSaur17}%
\begin{APACrefauthors}%
Hartkorn, O.%
\BCBT {}\ \BBA {} Saur, J.%
\end{APACrefauthors}%
\unskip\
\newblock
\APACrefYearMonthDay{2017}{}{}.
\newblock
{\BBOQ}\APACrefatitle {{Induction signals from Callisto's ionosphere and their
  implications on a possible subsurface ocean}} {{Induction signals from
  Callisto's ionosphere and their implications on a possible subsurface
  ocean}}.{\BBCQ}
\newblock
\APACjournalVolNumPages{J. Geophys. Res.}{122}{11}{}.
\PrintBackRefs{\CurrentBib}

\bibitem [\protect \citeauthoryear {%
Heimpel%
, Aurnou%
\BCBL {}\ \BBA {} Wicht%
}{%
Heimpel%
\ \protect \BOthers {.}}{%
{\protect \APACyear {2005}}%
}]{%
HeimpelEA05}
\APACinsertmetastar {%
HeimpelEA05}%
\begin{APACrefauthors}%
Heimpel, M\BPBI H.%
, Aurnou, J\BPBI M.%
\BCBL {}\ \BBA {} Wicht, J.%
\end{APACrefauthors}%
\unskip\
\newblock
\APACrefYearMonthDay{2005}{}{}.
\newblock
{\BBOQ}\APACrefatitle {{Simulation of equatorial and high-latitude jets on
  {J}upiter in a deep convection model}} {{Simulation of equatorial and
  high-latitude jets on {J}upiter in a deep convection model}}.{\BBCQ}
\newblock
\APACjournalVolNumPages{Nature}{483}{}{193--196}.
\PrintBackRefs{\CurrentBib}

\bibitem [\protect \citeauthoryear {%
Heimpel%
, Gastine%
\BCBL {}\ \BBA {} Wicht%
}{%
Heimpel%
\ \protect \BOthers {.}}{%
{\protect \APACyear {2015}}%
}]{%
HeimpelEA15}
\APACinsertmetastar {%
HeimpelEA15}%
\begin{APACrefauthors}%
Heimpel, M\BPBI H.%
, Gastine, T.%
\BCBL {}\ \BBA {} Wicht, J.%
\end{APACrefauthors}%
\unskip\
\newblock
\APACrefYearMonthDay{2015}{}{}.
\newblock
{\BBOQ}\APACrefatitle {{Simulation of deep-seated zonal jets and shallow
  vortices in gas giant atmospheres}} {{Simulation of deep-seated zonal jets
  and shallow vortices in gas giant atmospheres}}.{\BBCQ}
\newblock
\APACjournalVolNumPages{Nature Geosci.}{9}{1}{19}.
\PrintBackRefs{\CurrentBib}

\bibitem [\protect \citeauthoryear {%
Hemingway%
, Nimmo%
, Zebker%
\BCBL {}\ \BBA {} Iess%
}{%
Hemingway%
\ \protect \BOthers {.}}{%
{\protect \APACyear {2013}}%
}]{%
HemingwayEA13}
\APACinsertmetastar {%
HemingwayEA13}%
\begin{APACrefauthors}%
Hemingway, D.%
, Nimmo, F.%
, Zebker, H.%
\BCBL {}\ \BBA {} Iess, L.%
\end{APACrefauthors}%
\unskip\
\newblock
\APACrefYearMonthDay{2013}{}{}.
\newblock
{\BBOQ}\APACrefatitle {A rigid and weathered ice shell on {T}itan} {A rigid and
  weathered ice shell on {T}itan}.{\BBCQ}
\newblock
\APACjournalVolNumPages{Nature}{500}{}{550--552}.
\newblock
\begin{APACrefDOI} \doi{10.1038/nature12400} \end{APACrefDOI}
\PrintBackRefs{\CurrentBib}

\bibitem [\protect \citeauthoryear {%
Hori%
, Wicht%
\BCBL {}\ \BBA {} Christensen%
}{%
Hori%
\ \protect \BOthers {.}}{%
{\protect \APACyear {2012}}%
}]{%
HoriEA12}
\APACinsertmetastar {%
HoriEA12}%
\begin{APACrefauthors}%
Hori, K.%
, Wicht, J.%
\BCBL {}\ \BBA {} Christensen, U\BPBI R.%
\end{APACrefauthors}%
\unskip\
\newblock
\APACrefYearMonthDay{2012}{}{}.
\newblock
{\BBOQ}\APACrefatitle {The influence of thermo-compositional boundary
  conditions on convection and dynamos in a rotating spherical shell} {The
  influence of thermo-compositional boundary conditions on convection and
  dynamos in a rotating spherical shell}.{\BBCQ}
\newblock
\APACjournalVolNumPages{Phys. Earth Planet. Int.}{196--197}{}{32--48}.
\PrintBackRefs{\CurrentBib}

\bibitem [\protect \citeauthoryear {%
Jansen%
}{%
Jansen%
}{%
{\protect \APACyear {2016}}%
}]{%
Jansen16}
\APACinsertmetastar {%
Jansen16}%
\begin{APACrefauthors}%
Jansen, M\BPBI F.%
\end{APACrefauthors}%
\unskip\
\newblock
\APACrefYearMonthDay{2016}{}{}.
\newblock
{\BBOQ}\APACrefatitle {{The turbulent circulation of a Snowball Earth ocean}}
  {{The turbulent circulation of a Snowball Earth ocean}}.{\BBCQ}
\newblock
\APACjournalVolNumPages{J. Phys. Oceanogr.}{46}{6}{1917--1933}.
\PrintBackRefs{\CurrentBib}

\bibitem [\protect \citeauthoryear {%
Johnston%
\ \BBA {} Doering%
}{%
Johnston%
\ \BBA {} Doering%
}{%
{\protect \APACyear {2009}}%
}]{%
JohnstonDoering09}
\APACinsertmetastar {%
JohnstonDoering09}%
\begin{APACrefauthors}%
Johnston, H.%
\BCBT {}\ \BBA {} Doering, C\BPBI R.%
\end{APACrefauthors}%
\unskip\
\newblock
\APACrefYearMonthDay{2009}{}{}.
\newblock
{\BBOQ}\APACrefatitle {Comparison of turbulent thermal convection between
  conditions of constant temperature and constant flux} {Comparison of
  turbulent thermal convection between conditions of constant temperature and
  constant flux}.{\BBCQ}
\newblock
\APACjournalVolNumPages{Phys. Rev. Lett.}{102}{6}{064501}.
\PrintBackRefs{\CurrentBib}

\bibitem [\protect \citeauthoryear {%
Jones%
}{%
Jones%
}{%
{\protect \APACyear {2015}}%
}]{%
JonesTOG_2015}
\APACinsertmetastar {%
JonesTOG_2015}%
\begin{APACrefauthors}%
Jones, C\BPBI A.%
\end{APACrefauthors}%
\unskip\
\newblock
\APACrefYearMonthDay{2015}{}{}.
\newblock
{\BBOQ}\APACrefatitle {Thermal and compositional convection in the outer core}
  {Thermal and compositional convection in the outer core}.{\BBCQ}
\newblock
\BIn{} \APACrefbtitle {Treatise on Geophysics} {Treatise on geophysics}\
  (\BVOL~8, \BPGS\ 115--159).
\newblock
\APACaddressPublisher{}{Elsevier}.
\PrintBackRefs{\CurrentBib}

\bibitem [\protect \citeauthoryear {%
Julien%
, Knobloch%
, Rubio%
\BCBL {}\ \BBA {} Vasil%
}{%
Julien%
\ \protect \BOthers {.}}{%
{\protect \APACyear {2012}}%
}]{%
JulienEA12_PRL}
\APACinsertmetastar {%
JulienEA12_PRL}%
\begin{APACrefauthors}%
Julien, K.%
, Knobloch, E.%
, Rubio, A\BPBI M.%
\BCBL {}\ \BBA {} Vasil, G\BPBI M.%
\end{APACrefauthors}%
\unskip\
\newblock
\APACrefYearMonthDay{2012}{}{}.
\newblock
{\BBOQ}\APACrefatitle {{Heat transport in low-Rossby-number Rayleigh-B\'enard
  convection}} {{Heat transport in low-Rossby-number Rayleigh-B\'enard
  convection}}.{\BBCQ}
\newblock
\APACjournalVolNumPages{Phys. Rev. Lett.}{109}{}{254503}.
\PrintBackRefs{\CurrentBib}

\bibitem [\protect \citeauthoryear {%
Kalousov{\'a}%
, Schroeder%
\BCBL {}\ \BBA {} Soderlund%
}{%
Kalousov{\'a}%
\ \protect \BOthers {.}}{%
{\protect \APACyear {2017}}%
}]{%
KalousovaEA17}
\APACinsertmetastar {%
KalousovaEA17}%
\begin{APACrefauthors}%
Kalousov{\'a}, K.%
, Schroeder, D\BPBI M.%
\BCBL {}\ \BBA {} Soderlund, K\BPBI M.%
\end{APACrefauthors}%
\unskip\
\newblock
\APACrefYearMonthDay{2017}{}{}.
\newblock
{\BBOQ}\APACrefatitle {{Radar attenuation in Europa's ice shell: Obstacles and
  opportunities for constraining the shell thickness and its thermal
  structure}} {{Radar attenuation in Europa's ice shell: Obstacles and
  opportunities for constraining the shell thickness and its thermal
  structure}}.{\BBCQ}
\newblock
\APACjournalVolNumPages{J. Geophys. Res. Planets}{122}{3}{524--545}.
\PrintBackRefs{\CurrentBib}

\bibitem [\protect \citeauthoryear {%
Kuang%
\ \BBA {} Bloxham%
}{%
Kuang%
\ \BBA {} Bloxham%
}{%
{\protect \APACyear {1997}}%
}]{%
KuangBloxham97}
\APACinsertmetastar {%
KuangBloxham97}%
\begin{APACrefauthors}%
Kuang, W.%
\BCBT {}\ \BBA {} Bloxham, J.%
\end{APACrefauthors}%
\unskip\
\newblock
\APACrefYearMonthDay{1997}{}{}.
\newblock
{\BBOQ}\APACrefatitle {An Earth-like numerical dynamo model} {An earth-like
  numerical dynamo model}.{\BBCQ}
\newblock
\APACjournalVolNumPages{Nature}{389}{}{371--374}.
\PrintBackRefs{\CurrentBib}

\bibitem [\protect \citeauthoryear {%
Kvorka%
, Cadek%
, Tobie%
\BCBL {}\ \BBA {} Choblet%
}{%
Kvorka%
\ \protect \BOthers {.}}{%
{\protect \APACyear {2018}}%
}]{%
KvorkaEA18}
\APACinsertmetastar {%
KvorkaEA18}%
\begin{APACrefauthors}%
Kvorka, J.%
, Cadek, C.%
, Tobie, G.%
\BCBL {}\ \BBA {} Choblet, G.%
\end{APACrefauthors}%
\unskip\
\newblock
\APACrefYearMonthDay{2018}{}{}.
\newblock
{\BBOQ}\APACrefatitle {{Does Titan's long-wavelength topography contain
  information about subsurface ocean dynamics?}} {{Does Titan's long-wavelength
  topography contain information about subsurface ocean dynamics?}}{\BBCQ}
\newblock
\APACjournalVolNumPages{Icarus}{310}{}{149--164}.
\PrintBackRefs{\CurrentBib}

\bibitem [\protect \citeauthoryear {%
Le~Bars%
, Cebron%
\BCBL {}\ \BBA {} Le~Gal%
}{%
Le~Bars%
\ \protect \BOthers {.}}{%
{\protect \APACyear {2015}}%
}]{%
LeBarsEA15}
\APACinsertmetastar {%
LeBarsEA15}%
\begin{APACrefauthors}%
Le~Bars, M.%
, Cebron, D.%
\BCBL {}\ \BBA {} Le~Gal, P.%
\end{APACrefauthors}%
\unskip\
\newblock
\APACrefYearMonthDay{2015}{}{}.
\newblock
{\BBOQ}\APACrefatitle {Flows driven by libration, precession, and tides} {Flows
  driven by libration, precession, and tides}.{\BBCQ}
\newblock
\APACjournalVolNumPages{Annu. Rev. Fluid Mech.}{47}{}{163--193}.
\PrintBackRefs{\CurrentBib}

\bibitem [\protect \citeauthoryear {%
Lefevre%
, Tobie%
, Choblet%
\BCBL {}\ \BBA {} {\v{C}}adek%
}{%
Lefevre%
\ \protect \BOthers {.}}{%
{\protect \APACyear {2014}}%
}]{%
LefevreEA14}
\APACinsertmetastar {%
LefevreEA14}%
\begin{APACrefauthors}%
Lefevre, A.%
, Tobie, G.%
, Choblet, G.%
\BCBL {}\ \BBA {} {\v{C}}adek, O.%
\end{APACrefauthors}%
\unskip\
\newblock
\APACrefYearMonthDay{2014}{}{}.
\newblock
{\BBOQ}\APACrefatitle {Structure and dynamics of {T}itan's outer icy shell
  constrained from {C}assini data} {Structure and dynamics of {T}itan's outer
  icy shell constrained from {C}assini data}.{\BBCQ}
\newblock
\APACjournalVolNumPages{Icarus}{237}{}{16 -- 28}.
\newblock
\begin{APACrefDOI} \doi{10.1016/j.icarus.2014.04.006} \end{APACrefDOI}
\PrintBackRefs{\CurrentBib}

\bibitem [\protect \citeauthoryear {%
Lemasquerier%
\ \protect \BOthers {.}}{%
Lemasquerier%
\ \protect \BOthers {.}}{%
{\protect \APACyear {2017}}%
}]{%
LemasquerierEA17}
\APACinsertmetastar {%
LemasquerierEA17}%
\begin{APACrefauthors}%
Lemasquerier, D.%
, Grannan, A\BPBI M.%
, Vidal, J.%
, C{\'e}bron, D.%
, Favier, B.%
, Le~Bars, M.%
\BCBL {}\ \BBA {} Aurnou, J\BPBI M.%
\end{APACrefauthors}%
\unskip\
\newblock
\APACrefYearMonthDay{2017}{}{}.
\newblock
{\BBOQ}\APACrefatitle {Libration-driven flows in ellipsoidal shells}
  {Libration-driven flows in ellipsoidal shells}.{\BBCQ}
\newblock
\APACjournalVolNumPages{J. Geophys. Res. Planets}{122}{9}{1926--1950}.
\PrintBackRefs{\CurrentBib}

\bibitem [\protect \citeauthoryear {%
Leonard%
, Patthoff%
, Senske%
\BCBL {}\ \BBA {} Collins%
}{%
Leonard%
\ \protect \BOthers {.}}{%
{\protect \APACyear {2018}}%
}]{%
LeonardEA18}
\APACinsertmetastar {%
LeonardEA18}%
\begin{APACrefauthors}%
Leonard, E\BPBI J.%
, Patthoff, D\BPBI A.%
, Senske, D\BPBI A.%
\BCBL {}\ \BBA {} Collins, G\BPBI C.%
\end{APACrefauthors}%
\unskip\
\newblock
\APACrefYearMonthDay{2018}{}{}.
\newblock
{\BBOQ}\APACrefatitle {{The Europa Global Geologic Map}} {{The Europa Global
  Geologic Map}}.{\BBCQ}
\newblock
\APACjournalVolNumPages{LPI Contributions}{2066}{}{}.
\PrintBackRefs{\CurrentBib}

\bibitem [\protect \citeauthoryear {%
Lewis%
\ \BBA {} Perkin%
}{%
Lewis%
\ \BBA {} Perkin%
}{%
{\protect \APACyear {1986}}%
}]{%
LewisPerkin86}
\APACinsertmetastar {%
LewisPerkin86}%
\begin{APACrefauthors}%
Lewis, E\BPBI L.%
\BCBT {}\ \BBA {} Perkin, R\BPBI G.%
\end{APACrefauthors}%
\unskip\
\newblock
\APACrefYearMonthDay{1986}{}{}.
\newblock
{\BBOQ}\APACrefatitle {{Ice pumps and their rates}} {{Ice pumps and their
  rates}}.{\BBCQ}
\newblock
\APACjournalVolNumPages{J. Geophys. Res.}{91}{}{11756--11762}.
\PrintBackRefs{\CurrentBib}

\bibitem [\protect \citeauthoryear {%
Lucchita%
}{%
Lucchita%
}{%
{\protect \APACyear {1980}}%
}]{%
Lucchita80}
\APACinsertmetastar {%
Lucchita80}%
\begin{APACrefauthors}%
Lucchita, B\BPBI K.%
\end{APACrefauthors}%
\unskip\
\newblock
\APACrefYearMonthDay{1980}{}{}.
\newblock
{\BBOQ}\APACrefatitle {{Grooved terrain on Ganymede}} {{Grooved terrain on
  Ganymede}}.{\BBCQ}
\newblock
\APACjournalVolNumPages{Icarus}{44}{2}{481--501}.
\PrintBackRefs{\CurrentBib}

\bibitem [\protect \citeauthoryear {%
Lunine%
}{%
Lunine%
}{%
{\protect \APACyear {2017}}%
}]{%
Lunine17}
\APACinsertmetastar {%
Lunine17}%
\begin{APACrefauthors}%
Lunine, J\BPBI I.%
\end{APACrefauthors}%
\unskip\
\newblock
\APACrefYearMonthDay{2017}{}{}.
\newblock
{\BBOQ}\APACrefatitle {Ocean worlds exploration} {Ocean worlds
  exploration}.{\BBCQ}
\newblock
\APACjournalVolNumPages{Acta Astronautica}{131}{}{123--130}.
\PrintBackRefs{\CurrentBib}

\bibitem [\protect \citeauthoryear {%
Manneville%
\ \BBA {} Olson%
}{%
Manneville%
\ \BBA {} Olson%
}{%
{\protect \APACyear {1996}}%
}]{%
MannevilleOlson96}
\APACinsertmetastar {%
MannevilleOlson96}%
\begin{APACrefauthors}%
Manneville, J\BHBI B.%
\BCBT {}\ \BBA {} Olson, P\BPBI L.%
\end{APACrefauthors}%
\unskip\
\newblock
\APACrefYearMonthDay{1996}{}{}.
\newblock
{\BBOQ}\APACrefatitle {{Banded convection in rotating fluid spheres and the
  circulation of the Jovian atmosphere}} {{Banded convection in rotating fluid
  spheres and the circulation of the Jovian atmosphere}}.{\BBCQ}
\newblock
\APACjournalVolNumPages{Icarus}{122}{2}{242--250}.
\PrintBackRefs{\CurrentBib}

\bibitem [\protect \citeauthoryear {%
McCord%
, Hansen%
\BCBL {}\ \BBA {} Hibbitts%
}{%
McCord%
\ \protect \BOthers {.}}{%
{\protect \APACyear {2001}}%
}]{%
McCordEA01}
\APACinsertmetastar {%
McCordEA01}%
\begin{APACrefauthors}%
McCord, T\BPBI B.%
, Hansen, G\BPBI B.%
\BCBL {}\ \BBA {} Hibbitts, C\BPBI A.%
\end{APACrefauthors}%
\unskip\
\newblock
\APACrefYearMonthDay{2001}{}{}.
\newblock
{\BBOQ}\APACrefatitle {{Hydrated salt minerals on Ganymede's surface: evidence
  of an ocean below}} {{Hydrated salt minerals on Ganymede's surface: evidence
  of an ocean below}}.{\BBCQ}
\newblock
\APACjournalVolNumPages{Science}{292}{5521}{1523--1525}.
\PrintBackRefs{\CurrentBib}

\bibitem [\protect \citeauthoryear {%
Melosh%
, Ekholm%
, Showman%
\BCBL {}\ \BBA {} Lorenz%
}{%
Melosh%
\ \protect \BOthers {.}}{%
{\protect \APACyear {2004}}%
}]{%
MeloshEA04}
\APACinsertmetastar {%
MeloshEA04}%
\begin{APACrefauthors}%
Melosh, H\BPBI J.%
, Ekholm, A\BPBI G.%
, Showman, A\BPBI P.%
\BCBL {}\ \BBA {} Lorenz, R\BPBI D.%
\end{APACrefauthors}%
\unskip\
\newblock
\APACrefYearMonthDay{2004}{}{}.
\newblock
{\BBOQ}\APACrefatitle {The temperature of {E}uropa's subsurface water ocean}
  {The temperature of {E}uropa's subsurface water ocean}.{\BBCQ}
\newblock
\APACjournalVolNumPages{Icarus}{168}{}{498--502}.
\PrintBackRefs{\CurrentBib}

\bibitem [\protect \citeauthoryear {%
Miquel%
, Xie%
, Featherstone%
, Julien%
\BCBL {}\ \BBA {} Knobloch%
}{%
Miquel%
\ \protect \BOthers {.}}{%
{\protect \APACyear {2018}}%
}]{%
MiquelEA18}
\APACinsertmetastar {%
MiquelEA18}%
\begin{APACrefauthors}%
Miquel, B.%
, Xie, J\BHBI H.%
, Featherstone, N.%
, Julien, K.%
\BCBL {}\ \BBA {} Knobloch, E.%
\end{APACrefauthors}%
\unskip\
\newblock
\APACrefYearMonthDay{2018}{}{}.
\newblock
{\BBOQ}\APACrefatitle {Equatorially trapped convection in a rapidly rotating
  shallow shell} {Equatorially trapped convection in a rapidly rotating shallow
  shell}.{\BBCQ}
\newblock
\APACjournalVolNumPages{Phys. Rev. Fluids}{3}{5}{053801}.
\PrintBackRefs{\CurrentBib}

\bibitem [\protect \citeauthoryear {%
Mitri%
\ \protect \BOthers {.}}{%
Mitri%
\ \protect \BOthers {.}}{%
{\protect \APACyear {2014}}%
}]{%
MitriEA14}
\APACinsertmetastar {%
MitriEA14}%
\begin{APACrefauthors}%
Mitri, G.%
, Meriggiola, R.%
, Hayes, A.%
, Lefevre, A.%
, Tobie, G.%
, Genova, A.%
\BDBL {}Zebker, H.%
\end{APACrefauthors}%
\unskip\
\newblock
\APACrefYearMonthDay{2014}{}{}.
\newblock
{\BBOQ}\APACrefatitle {{Shape, topography, gravity anomalies and tidal
  deformation of Titan}} {{Shape, topography, gravity anomalies and tidal
  deformation of Titan}}.{\BBCQ}
\newblock
\APACjournalVolNumPages{Icarus}{236}{}{169--177}.
\PrintBackRefs{\CurrentBib}

\bibitem [\protect \citeauthoryear {%
Moore%
, Reid%
\BCBL {}\ \BBA {} Kipfstuhl%
}{%
Moore%
\ \protect \BOthers {.}}{%
{\protect \APACyear {1994}}%
}]{%
MooreEA94}
\APACinsertmetastar {%
MooreEA94}%
\begin{APACrefauthors}%
Moore, J\BPBI C.%
, Reid, A\BPBI P.%
\BCBL {}\ \BBA {} Kipfstuhl, J.%
\end{APACrefauthors}%
\unskip\
\newblock
\APACrefYearMonthDay{1994}{}{}.
\newblock
{\BBOQ}\APACrefatitle {Microphysical and electrical properties of marine ice
  and its relationship to meteoric and sea ice} {Microphysical and electrical
  properties of marine ice and its relationship to meteoric and sea
  ice}.{\BBCQ}
\newblock
\APACjournalVolNumPages{J. Geophys. Res.}{99}{}{5171--5180}.
\PrintBackRefs{\CurrentBib}

\bibitem [\protect \citeauthoryear {%
Mound%
\ \BBA {} Davies%
}{%
Mound%
\ \BBA {} Davies%
}{%
{\protect \APACyear {2017}}%
}]{%
MoundDavies17}
\APACinsertmetastar {%
MoundDavies17}%
\begin{APACrefauthors}%
Mound, J\BPBI E.%
\BCBT {}\ \BBA {} Davies, C\BPBI J.%
\end{APACrefauthors}%
\unskip\
\newblock
\APACrefYearMonthDay{2017}{}{}.
\newblock
{\BBOQ}\APACrefatitle {Heat transfer in rapidly rotating convection with
  heterogeneous thermal boundary conditions} {Heat transfer in rapidly rotating
  convection with heterogeneous thermal boundary conditions}.{\BBCQ}
\newblock
\APACjournalVolNumPages{J. Fluid Mech.}{828}{}{601--629}.
\PrintBackRefs{\CurrentBib}

\bibitem [\protect \citeauthoryear {%
Nayar%
, Sharqawy%
\BCBL {}\ \BBA {} Banchik%
}{%
Nayar%
\ \protect \BOthers {.}}{%
{\protect \APACyear {2016}}%
}]{%
NayarEA16}
\APACinsertmetastar {%
NayarEA16}%
\begin{APACrefauthors}%
Nayar, K\BPBI G.%
, Sharqawy, M\BPBI H.%
\BCBL {}\ \BBA {} Banchik, L\BPBI D.%
\end{APACrefauthors}%
\unskip\
\newblock
\APACrefYearMonthDay{2016}{}{}.
\newblock
{\BBOQ}\APACrefatitle {Thermophysical properties of seawater: a review and new
  correlations that include pressure dependence} {Thermophysical properties of
  seawater: a review and new correlations that include pressure
  dependence}.{\BBCQ}
\newblock
\APACjournalVolNumPages{Desalination}{390}{}{1--24}.
\PrintBackRefs{\CurrentBib}

\bibitem [\protect \citeauthoryear {%
Nimmo%
}{%
Nimmo%
}{%
{\protect \APACyear {2004}}%
}]{%
Nimmo04}
\APACinsertmetastar {%
Nimmo04}%
\begin{APACrefauthors}%
Nimmo, F.%
\end{APACrefauthors}%
\unskip\
\newblock
\APACrefYearMonthDay{2004}{}{}.
\newblock
{\BBOQ}\APACrefatitle {{Non-newtonian topographic relaxation on Europa}}
  {{Non-newtonian topographic relaxation on Europa}}.{\BBCQ}
\newblock
\APACjournalVolNumPages{Icarus}{168}{}{205--208}.
\PrintBackRefs{\CurrentBib}

\bibitem [\protect \citeauthoryear {%
Nimmo%
\ \BBA {} Bills%
}{%
Nimmo%
\ \BBA {} Bills%
}{%
{\protect \APACyear {2010}}%
}]{%
NimmoBills10}
\APACinsertmetastar {%
NimmoBills10}%
\begin{APACrefauthors}%
Nimmo, F.%
\BCBT {}\ \BBA {} Bills, B\BPBI G.%
\end{APACrefauthors}%
\unskip\
\newblock
\APACrefYearMonthDay{2010}{}{}.
\newblock
{\BBOQ}\APACrefatitle {{Shell thickness variations and the long-wavelength
  topography of Titan}} {{Shell thickness variations and the long-wavelength
  topography of Titan}}.{\BBCQ}
\newblock
\APACjournalVolNumPages{Icarus}{208}{2}{896--904}.
\PrintBackRefs{\CurrentBib}

\bibitem [\protect \citeauthoryear {%
Nimmo%
\ \BBA {} Pappalardo%
}{%
Nimmo%
\ \BBA {} Pappalardo%
}{%
{\protect \APACyear {2016}}%
}]{%
NimmoPappalardo16}
\APACinsertmetastar {%
NimmoPappalardo16}%
\begin{APACrefauthors}%
Nimmo, F.%
\BCBT {}\ \BBA {} Pappalardo, R\BPBI T.%
\end{APACrefauthors}%
\unskip\
\newblock
\APACrefYearMonthDay{2016}{}{}.
\newblock
{\BBOQ}\APACrefatitle {Ocean worlds in the outer solar system} {Ocean worlds in
  the outer solar system}.{\BBCQ}
\newblock
\APACjournalVolNumPages{J. Geophys. Res. Planets}{121}{8}{1378--1399}.
\PrintBackRefs{\CurrentBib}

\bibitem [\protect \citeauthoryear {%
Nimmo%
, Thomas%
, Pappalardo%
\BCBL {}\ \BBA {} Moore%
}{%
Nimmo%
\ \protect \BOthers {.}}{%
{\protect \APACyear {2007}}%
}]{%
NimmoEA07}
\APACinsertmetastar {%
NimmoEA07}%
\begin{APACrefauthors}%
Nimmo, F.%
, Thomas, P\BPBI C.%
, Pappalardo, R\BPBI T.%
\BCBL {}\ \BBA {} Moore, W\BPBI B.%
\end{APACrefauthors}%
\unskip\
\newblock
\APACrefYearMonthDay{2007}{}{}.
\newblock
{\BBOQ}\APACrefatitle {{The global shape of Europa: Constraints on lateral
  shell thickness variations}} {{The global shape of Europa: Constraints on
  lateral shell thickness variations}}.{\BBCQ}
\newblock
\APACjournalVolNumPages{Icarus}{191}{}{183--192}.
\PrintBackRefs{\CurrentBib}

\bibitem [\protect \citeauthoryear {%
Palguta%
, Anderson%
, Schubert%
\BCBL {}\ \BBA {} Moore%
}{%
Palguta%
\ \protect \BOthers {.}}{%
{\protect \APACyear {2006}}%
}]{%
PalgutaEA06}
\APACinsertmetastar {%
PalgutaEA06}%
\begin{APACrefauthors}%
Palguta, J.%
, Anderson, J\BPBI D.%
, Schubert, G.%
\BCBL {}\ \BBA {} Moore, W\BPBI B.%
\end{APACrefauthors}%
\unskip\
\newblock
\APACrefYearMonthDay{2006}{}{}.
\newblock
{\BBOQ}\APACrefatitle {{Mass anomalies on Ganymede}} {{Mass anomalies on
  Ganymede}}.{\BBCQ}
\newblock
\APACjournalVolNumPages{Icarus}{180}{}{428--441}.
\PrintBackRefs{\CurrentBib}

\bibitem [\protect \citeauthoryear {%
Pappalardo%
\ \BBA {} Barr%
}{%
Pappalardo%
\ \BBA {} Barr%
}{%
{\protect \APACyear {2004}}%
}]{%
PappalardoBarr04}
\APACinsertmetastar {%
PappalardoBarr04}%
\begin{APACrefauthors}%
Pappalardo, R\BPBI T.%
\BCBT {}\ \BBA {} Barr, A\BPBI C.%
\end{APACrefauthors}%
\unskip\
\newblock
\APACrefYearMonthDay{2004}{}{}.
\newblock
{\BBOQ}\APACrefatitle {The origin of domes on {E}uropa: The role of thermally
  induced compositional diapirism} {The origin of domes on {E}uropa: The role
  of thermally induced compositional diapirism}.{\BBCQ}
\newblock
\APACjournalVolNumPages{Geophys. Res. Lett.}{31}{}{L01701}.
\PrintBackRefs{\CurrentBib}

\bibitem [\protect \citeauthoryear {%
Phillips%
\ \BBA {} Pappalardo%
}{%
Phillips%
\ \BBA {} Pappalardo%
}{%
{\protect \APACyear {2014}}%
}]{%
PhillipsPappalardo14}
\APACinsertmetastar {%
PhillipsPappalardo14}%
\begin{APACrefauthors}%
Phillips, C\BPBI B.%
\BCBT {}\ \BBA {} Pappalardo, R\BPBI T.%
\end{APACrefauthors}%
\unskip\
\newblock
\APACrefYearMonthDay{2014}{}{}.
\newblock
{\BBOQ}\APACrefatitle {{Europa Clipper mission concept: Exploring Jupiter's
  ocean moon}} {{Europa Clipper mission concept: Exploring Jupiter's ocean
  moon}}.{\BBCQ}
\newblock
\APACjournalVolNumPages{{Eos, Transactions American Geophysical
  Union}}{95}{20}{165--167}.
\PrintBackRefs{\CurrentBib}

\bibitem [\protect \citeauthoryear {%
Rovira-Navarro%
\ \protect \BOthers {.}}{%
Rovira-Navarro%
\ \protect \BOthers {.}}{%
{\protect \APACyear {2019}}%
}]{%
RoviraNavarroEA19}
\APACinsertmetastar {%
RoviraNavarroEA19}%
\begin{APACrefauthors}%
Rovira-Navarro, M.%
, Rieutord, M.%
, Gerkema, T.%
, Maas, L\BPBI R.%
, van~der Wal, W.%
\BCBL {}\ \BBA {} Vermeersen, B.%
\end{APACrefauthors}%
\unskip\
\newblock
\APACrefYearMonthDay{2019}{}{}.
\newblock
{\BBOQ}\APACrefatitle {{Do tidally-generated inertial waves heat the subsurface
  oceans of Europa and Enceladus?}} {{Do tidally-generated inertial waves heat
  the subsurface oceans of Europa and Enceladus?}}{\BBCQ}
\newblock
\APACjournalVolNumPages{Icarus}{321}{}{126--140}.
\PrintBackRefs{\CurrentBib}

\bibitem [\protect \citeauthoryear {%
Sakuraba%
\ \BBA {} Roberts%
}{%
Sakuraba%
\ \BBA {} Roberts%
}{%
{\protect \APACyear {2009}}%
}]{%
SakurabaRoberts09}
\APACinsertmetastar {%
SakurabaRoberts09}%
\begin{APACrefauthors}%
Sakuraba, M.%
\BCBT {}\ \BBA {} Roberts, P\BPBI H.%
\end{APACrefauthors}%
\unskip\
\newblock
\APACrefYearMonthDay{2009}{}{}.
\newblock
{\BBOQ}\APACrefatitle {{Generation of a strong magnetic field using uniform
  heat flux at the surface of the core}} {{Generation of a strong magnetic
  field using uniform heat flux at the surface of the core}}.{\BBCQ}
\newblock
\APACjournalVolNumPages{Nature Geosci.}{2}{}{802--805}.
\PrintBackRefs{\CurrentBib}

\bibitem [\protect \citeauthoryear {%
Schmidt%
, Blankenship%
, Patterson%
\BCBL {}\ \BBA {} Schenk%
}{%
Schmidt%
\ \protect \BOthers {.}}{%
{\protect \APACyear {2011}}%
}]{%
SchmidtEA11}
\APACinsertmetastar {%
SchmidtEA11}%
\begin{APACrefauthors}%
Schmidt, B\BPBI E.%
, Blankenship, D\BPBI D.%
, Patterson, G\BPBI W.%
\BCBL {}\ \BBA {} Schenk, P\BPBI M.%
\end{APACrefauthors}%
\unskip\
\newblock
\APACrefYearMonthDay{2011}{}{}.
\newblock
{\BBOQ}\APACrefatitle {Active formation of chaos terrain over shallow
  subsurface water on {E}uropa} {Active formation of chaos terrain over shallow
  subsurface water on {E}uropa}.{\BBCQ}
\newblock
\APACjournalVolNumPages{Nature}{479}{}{502--505}.
\PrintBackRefs{\CurrentBib}

\bibitem [\protect \citeauthoryear {%
Soderlund%
, Heimpel%
, King%
\BCBL {}\ \BBA {} Aurnou%
}{%
Soderlund%
\ \protect \BOthers {.}}{%
{\protect \APACyear {2013}}%
}]{%
SoderlundEA13}
\APACinsertmetastar {%
SoderlundEA13}%
\begin{APACrefauthors}%
Soderlund, K\BPBI M.%
, Heimpel, M\BPBI H.%
, King, E\BPBI M.%
\BCBL {}\ \BBA {} Aurnou, J\BPBI M.%
\end{APACrefauthors}%
\unskip\
\newblock
\APACrefYearMonthDay{2013}{}{}.
\newblock
{\BBOQ}\APACrefatitle {Turbulent models of ice giant internal dynamics:
  Dynamos, heat transfer, and zonal flows} {Turbulent models of ice giant
  internal dynamics: Dynamos, heat transfer, and zonal flows}.{\BBCQ}
\newblock
\APACjournalVolNumPages{Icarus}{224}{}{97--113}.
\PrintBackRefs{\CurrentBib}

\bibitem [\protect \citeauthoryear {%
Soderlund%
, Schmidt%
, Wicht%
\BCBL {}\ \BBA {} Blankenship%
}{%
Soderlund%
\ \protect \BOthers {.}}{%
{\protect \APACyear {2014}}%
}]{%
SoderlundEA14}
\APACinsertmetastar {%
SoderlundEA14}%
\begin{APACrefauthors}%
Soderlund, K\BPBI M.%
, Schmidt, B\BPBI E.%
, Wicht, J.%
\BCBL {}\ \BBA {} Blankenship, D\BPBI D.%
\end{APACrefauthors}%
\unskip\
\newblock
\APACrefYearMonthDay{2014}{}{}.
\newblock
{\BBOQ}\APACrefatitle {{Ocean-driven heating of Europa's icy shell at low
  latitudes}} {{Ocean-driven heating of Europa's icy shell at low
  latitudes}}.{\BBCQ}
\newblock
\APACjournalVolNumPages{Nature Geosci.}{7}{}{16-19}.
\PrintBackRefs{\CurrentBib}

\bibitem [\protect \citeauthoryear {%
Sotin%
, Head%
\BCBL {}\ \BBA {} Tobie%
}{%
Sotin%
\ \protect \BOthers {.}}{%
{\protect \APACyear {2002}}%
}]{%
SotinEA02}
\APACinsertmetastar {%
SotinEA02}%
\begin{APACrefauthors}%
Sotin, C.%
, Head, J\BPBI W.%
\BCBL {}\ \BBA {} Tobie, G.%
\end{APACrefauthors}%
\unskip\
\newblock
\APACrefYearMonthDay{2002}{}{}.
\newblock
{\BBOQ}\APACrefatitle {{Tidal heating of upwelling thermal plumes and the
  origin of lenticulae and chaos melting}} {{Tidal heating of upwelling thermal
  plumes and the origin of lenticulae and chaos melting}}.{\BBCQ}
\newblock
\APACjournalVolNumPages{Geophys. Res. Lett.}{29}{8}{1233,
  doi:10.1029/2001GL013884}.
\PrintBackRefs{\CurrentBib}

\bibitem [\protect \citeauthoryear {%
Travis%
, Palguta%
\BCBL {}\ \BBA {} Schubert%
}{%
Travis%
\ \protect \BOthers {.}}{%
{\protect \APACyear {2012}}%
}]{%
TravisEA12}
\APACinsertmetastar {%
TravisEA12}%
\begin{APACrefauthors}%
Travis, B\BPBI J.%
, Palguta, J.%
\BCBL {}\ \BBA {} Schubert, G.%
\end{APACrefauthors}%
\unskip\
\newblock
\APACrefYearMonthDay{2012}{}{}.
\newblock
{\BBOQ}\APACrefatitle {{A whole-moon thermal history model of Europa: Impact of
  hydrothermal circulation and salt transport}} {{A whole-moon thermal history
  model of Europa: Impact of hydrothermal circulation and salt
  transport}}.{\BBCQ}
\newblock
\APACjournalVolNumPages{Icarus}{218}{}{1006--1019}.
\PrintBackRefs{\CurrentBib}

\bibitem [\protect \citeauthoryear {%
Tyler%
}{%
Tyler%
}{%
{\protect \APACyear {2008}}%
}]{%
Tyler08}
\APACinsertmetastar {%
Tyler08}%
\begin{APACrefauthors}%
Tyler, R\BPBI H.%
\end{APACrefauthors}%
\unskip\
\newblock
\APACrefYearMonthDay{2008}{}{}.
\newblock
{\BBOQ}\APACrefatitle {Strong ocean tidal flow and heating on moons of the
  outer planets} {Strong ocean tidal flow and heating on moons of the outer
  planets}.{\BBCQ}
\newblock
\APACjournalVolNumPages{Nature}{456}{}{770--773}.
\PrintBackRefs{\CurrentBib}

\bibitem [\protect \citeauthoryear {%
Tyler%
}{%
Tyler%
}{%
{\protect \APACyear {2011}}%
}]{%
Tyler11}
\APACinsertmetastar {%
Tyler11}%
\begin{APACrefauthors}%
Tyler, R\BPBI H.%
\end{APACrefauthors}%
\unskip\
\newblock
\APACrefYearMonthDay{2011}{}{}.
\newblock
{\BBOQ}\APACrefatitle {{Magnetic remote sensing of Europa's ocean tides}}
  {{Magnetic remote sensing of Europa's ocean tides}}.{\BBCQ}
\newblock
\APACjournalVolNumPages{Icarus}{211}{}{906--908}.
\PrintBackRefs{\CurrentBib}

\bibitem [\protect \citeauthoryear {%
Vance%
}{%
Vance%
}{%
{\protect \APACyear {2017}}%
}]{%
Vance17}
\APACinsertmetastar {%
Vance17}%
\begin{APACrefauthors}%
Vance, S.%
\end{APACrefauthors}%
\unskip\
\newblock
\APACrefYearMonthDay{2017}{}{}.
\newblock
{\BBOQ}\APACrefatitle {vancesteven/planetprofile: Release for use in
  reproducing results submitted to Journal of Geophysical Research: Planets}
  {vancesteven/planetprofile: Release for use in reproducing results submitted
  to journal of geophysical research: Planets}.{\BBCQ}
\newblock
\APACjournalVolNumPages{Zenodo}{https://doi.org/10.5281/zenodo.844131}{}{}.
\PrintBackRefs{\CurrentBib}

\bibitem [\protect \citeauthoryear {%
Vance%
\ \BBA {} Brown%
}{%
Vance%
\ \BBA {} Brown%
}{%
{\protect \APACyear {2005}}%
}]{%
VanceBrown05}
\APACinsertmetastar {%
VanceBrown05}%
\begin{APACrefauthors}%
Vance, S.%
\BCBT {}\ \BBA {} Brown, J\BPBI M.%
\end{APACrefauthors}%
\unskip\
\newblock
\APACrefYearMonthDay{2005}{}{}.
\newblock
{\BBOQ}\APACrefatitle {Layering and double-diffusion style convection in
  {E}uropa's ocean} {Layering and double-diffusion style convection in
  {E}uropa's ocean}.{\BBCQ}
\newblock
\APACjournalVolNumPages{Icarus}{177}{}{506--514}.
\PrintBackRefs{\CurrentBib}

\bibitem [\protect \citeauthoryear {%
Vance%
\ \BBA {} Goodman%
}{%
Vance%
\ \BBA {} Goodman%
}{%
{\protect \APACyear {2009}}%
}]{%
VanceGoodman09}
\APACinsertmetastar {%
VanceGoodman09}%
\begin{APACrefauthors}%
Vance, S.%
\BCBT {}\ \BBA {} Goodman, J\BPBI C.%
\end{APACrefauthors}%
\unskip\
\newblock
\APACrefYearMonthDay{2009}{}{}.
\newblock
{\BBOQ}\APACrefatitle {Oceanography of an ice-covered moon} {Oceanography of an
  ice-covered moon}.{\BBCQ}
\newblock
\BIn{} R\BPBI T.~Pappalardo, W\BPBI B.~McKinnon\BCBL {}\ \BBA {} K\BPBI
  K.~Khurana\ (\BEDS), \APACrefbtitle {{E}uropa} {{E}uropa}\ (\BPG~459-482).
\newblock
\APACaddressPublisher{}{Tucson: University of Arizona Press}.
\PrintBackRefs{\CurrentBib}

\bibitem [\protect \citeauthoryear {%
Vance%
\ \protect \BOthers {.}}{%
Vance%
\ \protect \BOthers {.}}{%
{\protect \APACyear {2018}}%
}]{%
VanceEA18}
\APACinsertmetastar {%
VanceEA18}%
\begin{APACrefauthors}%
Vance, S.%
, Panning, M\BPBI P.%
, Stahler, S.%
, Cammarano, F.%
, Bills, B\BPBI G.%
, Tobie, G.%
\BDBL {}Banerdt, B.%
\end{APACrefauthors}%
\unskip\
\newblock
\APACrefYearMonthDay{2018}{}{}.
\newblock
{\BBOQ}\APACrefatitle {{Geophysical investigations of habitability in
  ice-covered ocean worlds}} {{Geophysical investigations of habitability in
  ice-covered ocean worlds}}.{\BBCQ}
\newblock
\APACjournalVolNumPages{J. Geophys. Res.}{123}{}{180--205}.
\PrintBackRefs{\CurrentBib}

\bibitem [\protect \citeauthoryear {%
Wicht%
}{%
Wicht%
}{%
{\protect \APACyear {2002}}%
}]{%
Wicht02}
\APACinsertmetastar {%
Wicht02}%
\begin{APACrefauthors}%
Wicht, J.%
\end{APACrefauthors}%
\unskip\
\newblock
\APACrefYearMonthDay{2002}{}{}.
\newblock
{\BBOQ}\APACrefatitle {{Inner-core conductivity in numerical dynamo
  simulations}} {{Inner-core conductivity in numerical dynamo
  simulations}}.{\BBCQ}
\newblock
\APACjournalVolNumPages{Phys. Earth Planet. Int.}{132}{}{281--302}.
\PrintBackRefs{\CurrentBib}

\bibitem [\protect \citeauthoryear {%
Wilson%
\ \BBA {} Kerswell%
}{%
Wilson%
\ \BBA {} Kerswell%
}{%
{\protect \APACyear {2018}}%
}]{%
WilsonKerswell18}
\APACinsertmetastar {%
WilsonKerswell18}%
\begin{APACrefauthors}%
Wilson, A.%
\BCBT {}\ \BBA {} Kerswell, R\BPBI R.%
\end{APACrefauthors}%
\unskip\
\newblock
\APACrefYearMonthDay{2018}{}{}.
\newblock
{\BBOQ}\APACrefatitle {{Can libration maintain Enceladus's ocean?}} {{Can
  libration maintain Enceladus's ocean?}}{\BBCQ}
\newblock
\APACjournalVolNumPages{Earth Planet. Sci. Lett.}{500}{}{41 -- 46}.
\newblock
\begin{APACrefDOI} \doi{10.1016/j.epsl.2018.08.012} \end{APACrefDOI}
\PrintBackRefs{\CurrentBib}

\bibitem [\protect \citeauthoryear {%
Zhang%
\ \BBA {} Schubert%
}{%
Zhang%
\ \BBA {} Schubert%
}{%
{\protect \APACyear {2000}}%
}]{%
ZhangSchubert00}
\APACinsertmetastar {%
ZhangSchubert00}%
\begin{APACrefauthors}%
Zhang, K.%
\BCBT {}\ \BBA {} Schubert, G.%
\end{APACrefauthors}%
\unskip\
\newblock
\APACrefYearMonthDay{2000}{}{}.
\newblock
{\BBOQ}\APACrefatitle {{Magnetohydrodynamics in rapidly rotating spherical
  systems}} {{Magnetohydrodynamics in rapidly rotating spherical
  systems}}.{\BBCQ}
\newblock
\APACjournalVolNumPages{Annu. Rev. Fluid Mech.}{32}{}{409--443}.
\PrintBackRefs{\CurrentBib}

\end{thebibliography}
%
% no need to specify bibliographystyle
%%%%%%%%%%%%%%%%%%%%%%%%%%%%%%%%%%%%%%%%%%%%%%%

% Please use ONLY \citet and \citep for reference citations.
% DO NOT use other cite commands (e.g., \cite, \citeyear, \nocite, \citealp, etc.).
%% Example \citet and \citep:
%  ...as shown by \citet{Boug10}, \citet{Buiz07}, \citet{Fra10},
%  \citet{Ghel00}, and \citet{Leit74}.

%  ...as shown by \citep{Boug10}, \citep{Buiz07}, \citep{Fra10},
%  \citep{Ghel00, Leit74}.

%  ...has been shown \citep [e.g.,][]{Boug10,Buiz07,Fra10}.

\begin{table}
\caption{Properties of icy satellite oceans in dimensional units and non-dimensional parameters. Kinematic viscosity from \cite{NayarEA16}, thermal diffusivity from \cite{AbramsonEA01}, interior model properties ($R_S, D_{Ih}, D_{ocean}, q$) from \cite{VanceEA18}, and thermodynamic properties ($\rho, C_p, \alpha$) at the mean ocean temperatures and pressures of the respective interior models from \cite{Vance17} for MgSO$_4$ (0 and 10 wt \%) and B. Journaux (personal communication) for seawater. Two outer ice shell thicknesses and three ocean compositions are considered for Enceladus and Europa, while three outer ice shell thicknesses and two ocean compositions are considered for Titan and Ganymede.} 
\centering
\begin{tabular}{lcccc}
\hline
                                            & Enceladus             & Titan                 & Europa                & Ganymede              \\
\hline
Gravitational acceleration, $g$ [m/s$^{2}$] & $0.1$                 & $1.4$                 & $1.3$                 & $1.4$                 \\ 
Rotation rate, $\Omega$ [s$^{-1}$] 			& $5.3 \times 10^{-5}$  & $4.6 \times 10^{-6}$  & $2.1 \times 10^{-5}$  & $1.0 \times 10^{-5}$  \\
Kinematic viscosity, $\nu$ [m$^2$/s] 		& $1.8 \times 10^{-6}$  & $1.8 \times 10^{-6}$  & $1.8 \times 10^{-6}$  & $1.8 \times 10^{-6}$  \\
Thermal diffusivity, $\kappa$ [m$^2$/s] 	& $1.4 \times 10^{-7}$  & $1.8 \times 10^{-7}$  & $1.6 \times 10^{-7}$  & $1.8 \times 10^{-7}$  \\
Satellite radius, $R_S$ [km]                & $252$                 & $2575$                & $1561$                & $2631$   \\    
Ice Ih thickness, $D_{Ih}$ [km] 	        &                       &                       &                       &                       \\
\hspace{5mm} {\it Water}                    & $51, 10$              & $141, 74, 50$         & $30, 5$               & $134, 70, 5$          \\
\hspace{5mm} {\it MgSO$_4$ 10 wt\%}         & $50, 10$              & $149, 86, 58$         & $30, 5$               & $157, 95, 26$         \\
\hspace{5mm} {\it Seawater}                 & $50, 10$              & $-$                   & $30, 5$               & $-$                   \\

Ocean thickness, $D_{ocean}$ [km] 			&                       &                       &                       &                       \\
\hspace{5mm} {\it Water}                    & $11, 53$              & $130, 369, 420$       & $97, 124$             & $119, 361, 518$        \\
\hspace{5mm} {\it MgSO$_4$ 10 wt\%}         & $13, 63$              & $91, 333, 403$        & $103, 131$            & $24, 287, 493$        \\
\hspace{5mm} {\it Seawater}                 & $12, 55$              & $-$                   & $99, 126$             & $-$                   \\

Heat flux, $q$ [mW/m$^{2}$] 				&                       &                       &                       &                       \\
\hspace{5mm} {\it Water}                    & $16, 81$              & $14, 18, 20$          & $24, 119$             & $16, 20, 107$         \\
\hspace{5mm} {\it MgSO$_4$ 10 wt\%}         & $16, 83$              & $14, 17, 19$          & $24, 123$             & $15, 18, 25$          \\
\hspace{5mm} {\it Seawater}                 & $16, 82$              & $-$                   & $23, 121$             & $-$                   \\

Density, $\rho$ [$10^3$ kg/m$^{3}$] 	    &                       &                       &                       &                       \\
\hspace{5mm} {\it Water}                    & $1.00, 1.00$          & $1.11, 1.14, 1.14$    & $1.04, 1.04$          & $1.11, 1.14, 1.14$    \\ 
\hspace{5mm} {\it MgSO$_4$ 10 wt\%}         & $1.11, 1.11$          & $1.20, 1.23, 1.24$    & $1.15, 1.14$          & $1.19, 1.23, 1.24$    \\ 
\hspace{5mm} {\it Seawater}                 & $1.02, 1.02$          & $-$                   & $1.07, 1.07$          & $-$                   \\

Specific heat capacity, $C_p$ [$10^3$ J/kg/K] &                     &                       &                       &                       \\
\hspace{5mm} {\it Water}                    & $4.2, 4.2$            & $3.0, 3.5, 3.6$       & $3.9, 3.9$           & $3.0, 3.5, 3.7$       \\
\hspace{5mm} {\it MgSO$_4$ 10 wt\%}         & $3.6, 3.7$            & $2.1, 2.5, 2.8$       & $3.3, 3.5$           & $2.1, 2.4, 3.0$       \\
\hspace{5mm} {\it Seawater}                 & $4.0, 4.0$            & $-$                   & $3.8, 3.8$           & $-$                   \\

Thermal expansivity, $\alpha$ [$10^{-4}$ K$^{-1}$] &                &                       &                       &                       \\
\hspace{5mm} {\it Water}                    & $-0.5, -0.5$          & $2.3, 4.0, 4.2$       & $1.9, 1.9$            & $2.2, 4.0, 4.4$       \\
\hspace{5mm} {\it MgSO$_4$ 10 wt\%}         & $1.2, 1.3$            & $0.4, 2.1, 2.7$       & $2.1, 2.3$            & $-0.1, 1.9, 3.2$      \\
\hspace{5mm} {\it Seawater}                 & $0.1, 0.1$            & $-$                   & $2.5, 2.5$            & $-$                   \\
\hline 
Prandtl number, $Pr = \nu / \kappa $ 				     & 13                       & 10                    & 11                    & 10 \\
Ekman number, $E=\nu / \Omega D^2$ 					     & $10^{-10} - 10^{-11}$    & $10^{-11} - 10^{-12}$ & $10^{-12}$            & $10^{-10} - 10^{-13}$ \\
Rayleigh number, $Ra=\alpha g \Delta T D^3 / \nu \kappa$ & $10^{16} - 10^{19}$      & $10^{19} - 10^{23}$   & $10^{20} - 10^{22}$   & $10^{20} - 10^{24}$ \\
Radius ratio, $\chi = \frac{(R_S - D_{Ih})}{(R_S - D_{Ih} - D_{ocean})}$  		    & $0.74 - 0.95$         & $0.83 - 0.96$         & $0.92 - 0.94$         & $0.80 - 0.99$ \\
\hline
\label{tab:physparam}
\end{tabular}
\end{table}
\clearpage

\clearpage

\nolinenumbers
\includepdf[pages=-,pagecommand={\thispagestyle{empty}}]{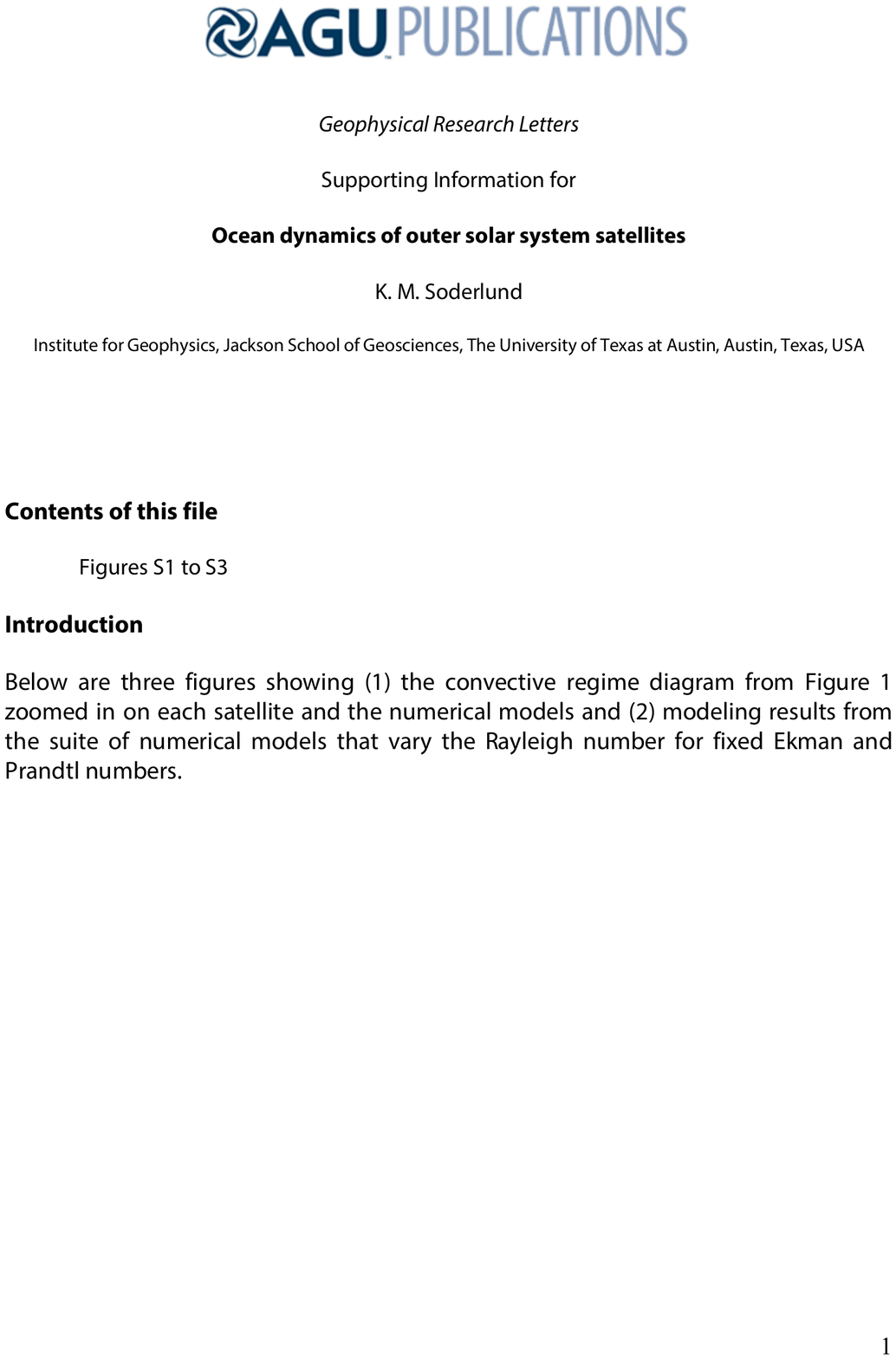}

\end{document}